\documentclass[12pt]{iopart}
\usepackage[T1]{fontenc}
\usepackage[latin9]{inputenc}
\usepackage{amsbsy}
\usepackage{amstext}
\usepackage{graphicx}
\PassOptionsToPackage{normalem}{ulem}
\usepackage{ulem}

\makeatletter
\def\ssr{\ref@jnl{Space~Sci.~Rev.}}     
\expandafter\let\csname equation*\endcsname\relax 
\expandafter\let\csname endequation*\endcsname\relax
\usepackage{iopams}
\usepackage{setstack}
\usepackage{aas_macros}

\usepackage[numbers, sort&compress]{natbib}

\def\newblock{\hskip .11em plus .33em minus .07em}

\newcommand{\eqref}[1]{(\ref{#1})}

\@ifundefined{showcaptionsetup}{}{%
 \PassOptionsToPackage{caption=false}{subfig}}
\usepackage{subfig}
\makeatother

\begin{document}
\global\long\def\g{\nabla}
\global\long\def\E{\boldsymbol{E}}
\global\long\def\i{\mathrm{i}}
\global\long\def\x{\boldsymbol{x}}
\global\long\def\B{\boldsymbol{B}}
\global\long\def\vv{\boldsymbol{v}}
\global\long\def\gv{\g_{v}}
\global\long\def\n{\delta n_{\i}}
\global\long\def\f{\delta f_{\i}}
\global\long\def\de{\delta\E}
\global\long\def\db{\delta\B}
\global\long\def\F{F_{0\i}}
\global\long\def\k{\boldsymbol{k}}
\global\long\def\d{\cdot}
\global\long\def\t{\times}
\global\long\def\kx{k_{\perp}}
\global\long\def\kp{k_{\|}}
\global\long\def\u{\boldsymbol{u}}
\global\long\def\sp{\sigma}
\global\long\def\oo{\infty}
\global\long\def\tpa{T_{0\i\|}}
\global\long\def\tp{T_{0\i\perp}}
\global\long\def\vtpa{v_{\mathrm{th}\i\|}}
\global\long\def\vtp{v_{\mathrm{th}\i\perp}}
\global\long\def\vp{v_{\perp}}
\global\long\def\vpa{v_{\|}}
\global\long\def\dby{\delta B_{y}}
\global\long\def\dbp{\delta B_{\|}}
\global\long\def\dbx{\delta B_{x}}
\global\long\def\c{c_{\mathrm{ref}}}
\global\long\def\kh{\hat{k}}
\global\long\def\kph{\hat{k}_{\|}}
\global\long\def\kxh{\hat{k}_{\perp}}
\global\long\def\vph{\hat{v}_{\perp}}
\global\long\def\vpah{\hat{v}_{\|}}
\global\long\def\Alf{\textrm{Alfvén}}
\global\long\def\oci{\Omega_{\mathrm{ci}}}
\global\long\def\oce{\Omega_{\mathrm{ce}}}
\global\long\def\va{v_{\mathrm{A}}}
\global\long\def\di{d_{\i}}
\global\long\def\m{m_{\i}}
\global\long\def\e{\mathrm{e}}
\global\long\def\p{\mathrm{p}}
\global\long\def\im{\mathrm{i}}

\title[A linear dispersion relation for the hybrid-kinetic model]{A linear dispersion relation for the hybrid kinetic-ion/fluid-electron
model of plasma physics}

\author{D~Told$^{1}$, J~Cookmeyer$^{1,2}$, P~Astfalk$^{3}$, F~Jenko$^{1}$}

\address{1) Department of Physics and Astronomy, University of California,
Los Angeles, CA 90095, USA}

\address{2) Haverford College, 370 Lancaster Avenue, Haverford, PA 19041,
USA}

\address{3) Max-Planck-Institut für Plasmaphysik, Boltzmannstr. 2, D-85748
Garching, Germany}

\submitto{\NJP}
\begin{abstract}
A dispersion relation for a commonly used hybrid model of plasma physics
is developed, which combines fully kinetic ions and a massless-electron
fluid description. Although this model and variations of it have been
used to describe plasma phenomena for about 40 years, to date there
exists no general dispersion relation to describe the linear wave
physics contained in the model. Previous efforts along these lines
are extended here to retain arbitrary wave propagation angles, temperature
anisotropy effects, as well as additional terms in the generalized
Ohm's law which determines the electric field. A numerical solver
for the dispersion relation is developed, and linear wave physics
is benchmarked against solutions of a full Vlasov-Maxwell dispersion
relation solver. This work opens the door to a more accurate interpretation
of existing and future wave and turbulence simulations using this
type of hybrid model. 
\end{abstract}
\maketitle

\section{Introduction}

Phenomena such as plasma instabilities, magnetic reconnection, turbulent
transport and dissipation of energy have been studied for several
decades through experiment and observation, and in a variety of systems
ranging from laboratory plasmas like linear devices  and tokamaks
 to natural plasmas like the solar corona, solar wind, or planetary
magnetospheres (e.g., Refs.~\cite{Gekelman99,Reme01,Shimada07,Marsch06,Bruno13}).
The plasmas permeating these systems are in many cases sufficiently
collisionless to support a host of kinetic effects, with strong influence
on the evolution of those systems. A comprehensive theoretical description
of such a plasma often proves to be very complex, as analytical solutions
of the kinetic equations are usually constrained to simplified cases
or linearized problems. Numerical methods must then be employed in
order to model more realistic situations. 

However, because of the intrinsic separation between ion and electron
spatiotemporal scales in a plasma, in many cases numerically simulating
the full Vlasov-Maxwell system of equations for all involved species
is also not feasible due to the significant computational expense
of evolving a six-dimensional system (3 spatial, 3 velocity-space
dimensions) across the diverse timescales that are involved. In practice,
many simplifications such as reduced spatial dimensionality, an artificially
reduced ion/electron mass ratio, or analytically reduced models such
as gyrokinetics \cite{Howes06,Howes11,Told15} are employed to make
such investigations tractable. 

An alternative and very common approach along these lines is to introduce
a hybrid model that applies the fully kinetic treatment only to the
ions, while describing the electrons within the framework of a simpler
fluid model. Historically, such models have been used for about 40
years \cite{Chodura75,Sgro76,Harned82,Winske85,Brecht88,Swift96,Winske03},
but surprisingly, to our knowledge there exists no general dispersion
relation to describe the linear wave physics within such a reduced
model. Limiting cases for perpendicular propagation \cite{Kazeminezhad92}
and for parametric instabilities of $\Alf$ waves propagating parallel
to the background magnetic field \cite{Araneda07} have been derived,
but the general case of oblique propagation has not been treated in
the literature. Existing simulation codes based on this model (see,
e.g., Refs.~\cite{Gargate07,Valentini07,Mueller11,Kempf13,Cheng13,Kunz14}
for recent efforts) are often benchmarked against frequencies (more
rarely against growth or damping rates) obtained from analytical or
full Vlasov-Maxwell dispersion relations.

In the present paper, we intend to fill this gap by deriving a dispersion
relation for the hybrid kinetic ion, massless electron model, variations
of which are being actively used in a number of simulation codes throughout
the space and astrophysics community. The existence of such a dispersion
relation will enable a more thorough and clear-cut validation of the
hybrid model itself, and also of simulations making use of that model.
For this purpose, we have developed the numerical dispersion solver
``HYDROS'', which is publicly available \cite{Hydros} and has already
been used to help interpret nonlinear simulation results in a recent
publication \cite{Cerri16}.

This paper is structured as follows: In Sec.~\ref{sec:Eq}, we describe
briefly the set of equations that will form the basis for our dispersion
relation. In the main part of the paper, Sec.~\ref{sec:deriv} is
devoted to the derivation of the dispersion relation, while in Sec.~\ref{sec:Hydros}
we describe the implementation of the ``HYDROS'' solver, together
with a variety of benchmark tests against the fully kinetic code DSHARK
\cite{Astfalk15}. Sec.~\ref{sec:conclusions}, finally, provides
a summary of this work.

\section{The hybrid-kinetic system of equations}

\label{sec:Eq}

In this section, we introduce the hybrid kinetic-ion/fluid-electron
(``hybrid-kinetic'' in the following) equations that will be used
throughout this paper. For our present purposes, we stick to a nonrelativistic,
low-frequency (i.e. $\omega\ll\Omega_{\mathrm{ce}}$) version of the
equations which assumes massless electrons and retains only a singly
charged ion species such that $n_{\i}=n_{\e}$. The system of equations
then consists of the ion Vlasov equation,
\begin{equation}
\frac{\partial f_{\i}}{\partial t}+\vv\cdot\g f_{\i}+\left[\frac{e}{m_{\i}}\left(\E+\frac{\vv\times\B}{c}\right)\right]\cdot\gv f_{\i}=0,\label{eq:Vl}
\end{equation}
and an Ohm's law which determines the electric field,
\begin{equation}
n_{\e}\E=-\frac{1}{c}n_{\i}\u_{\i}\times\B+\frac{1}{ce}\boldsymbol{j}\times\B-\frac{1}{e}\nabla P_{\e}+n_{\e}\eta\boldsymbol{j}\label{eq:Ohm}
\end{equation}
with $n_{i}\u_{\i}=\int\vv f_{\i}d^{3}v$, the resistivity $\eta$,
and the electron pressure gradient $\nabla P_{\e}=C\g n_{\e}^{\gamma}$.
The electromagnetic fields are further constrained by Faraday's law
\begin{equation}
\frac{\partial\B}{\partial t}=-c\g\times\E,\label{eq:Faraday}
\end{equation}
and the nonrelativistic version of Ampere's law
\begin{equation}
\g\t\B=\frac{4\pi}{c}\boldsymbol{j}\label{eq:amp}
\end{equation}
These equations contain the full ion kinetic physics including wave-particle
interactions such as Landau and transit-time damping, as well as cyclotron
resonances. The ion background distribution is assumed to be bi-Maxwellian,
enabling the occurrence of the basic ion-anisotropy driven instabilities
such as the firehose and mirror modes. 

Electrons, on the other hand, appear only as a neutralizing, massless
background species (where $C=n_{\e}^{1-\gamma}T_{0\e}$, and the choice
$\gamma=1$ or $5/3$ results in an isothermal or adiabatic electron
model for the electron pressure gradient term) and implicitly as the
carriers of the current $\boldsymbol{j}$. Interactions between waves
and electrons are not retained, so that, e.g., electron Landau damping
is absent.

\section{Derivation of the hybrid-kinetic Vlasov-Maxwell dispersion relation}

\label{sec:deriv}This section details the derivation of a dispersion
relation which completely describes the linear wave physics contained
in the system defined by Eqs.~(\ref{eq:Vl})--(\ref{eq:amp}). Here,
we will loosely follow the procedure employed before for perpendicularly
propagating waves in a hybrid-kinetic model in Ref.~\cite{Kazeminezhad92},
although from the outset we retain a general wave vector, as well
as additional terms in Ohm's law and an anisotropic background distribution.
We note that alternatively, it is possible to derive a hybrid-kinetic
dispersion relation using the dielectric tensor method described,
e.g., in Ref.~\cite{Stix62}. Since the dielectric is additive in
species, it is possible to combine the fully kinetic ion susceptibility
with that of a fluid electron model, derived, e.g. in Ref.~\cite{Swanson89}.

As a first step, we linearize Eqs.~(\ref{eq:Vl})--(\ref{eq:amp})
with respect to a static, homogeneous, background, according to the
rules $f_{\i}=F_{0\i}+\delta f_{\i}$, $n_{\i}=n_{0\i}+\delta n_{\i}$,
$\B=B_{0}\boldsymbol{z}+\delta\B$, and $\E=\delta\E$. The ion density
perturbation is defined as $\n=\int\f d^{3}v$. Furthermore, we introduce
a plane wave expansion for all perturbed quantities
\[
\delta A=\sum_{k}\delta\tilde{A}_{k}\exp\left(\im\left(\k\d\x-\omega t\right)\right).
\]
 For the purposes of this paper, we will assume the background ion
distribution to have the shape of a gyrotropic Maxwellian, retaining
the effects of an anisotropic temperature. The background distribution
can thus be written as 
\begin{equation}
F_{0\i}=n_{0\i}\left(\frac{m_{i}}{2\pi}\right)^{3/2}\left(\tpa T_{0\i\perp}^{2}\right)^{-1/2}\exp\left(-\frac{m_{\i}v_{\|}^{2}}{2\tpa}-\frac{m_{\i}\vp^{2}}{2\tp}\right).\label{eq:F0}
\end{equation}
Under such an assumption, we may choose the alignment of our coordinate
system such that, without loss of generality,
\[
\k=\left(\kx,0,\kp\right)^{T}.
\]
The divergence-free property of the magnetic field then reads 
\[
\k\cdot\db=0,
\]
so that the magnetic fluctuations in $x$ and parallel ($z$) direction
are related via 
\begin{equation}
\delta B_{x}=-\frac{\kp\delta B_{\|}}{\kx},\label{eq:dbx}
\end{equation}
and we need to describe only $\dby$ and $\dbp$ separately. Applying
the linearization rules to Eqs.~(\ref{eq:Vl})--(\ref{eq:Faraday}),
setting $n_{\e}=n_{\i}$, and using Eq.~(\ref{eq:amp}) to replace
the current yields
\begin{equation}
\fl\frac{\partial\f}{\partial t}+\vv\cdot\g\f+\left[\frac{e}{\m}\left(\de+\frac{\vv\times\db}{c}\right)\right]\cdot\gv\F+\frac{e}{\m}\left(\frac{\vv\times\B_{0}}{c}\right)\cdot\gv\f=0\label{eq:linVlasov}
\end{equation}
\begin{eqnarray}
\fl n_{0\i}\delta\E=-\frac{1}{c}n_{0\i}\u_{\i}\times\B_{0}+\frac{1}{4\pi e}\left(\g\times\delta\B\right)\times\B_{0}-\frac{C}{e}\gamma\g\n\left(n_{0\i}+\n\right)^{\gamma-1}\label{eq:linE}\\
+\frac{n_{0\i}\eta c}{4\pi}\g\times\delta\B\nonumber 
\end{eqnarray}
\begin{equation}
\frac{\partial\db}{\partial t}=-c\g\times\de\label{eq:linFar}
\end{equation}
Because we chose a static background, we may now write $\u_{\i}=\int\vv\f d^{3}v/n_{0\i}$.
In Eq.~(\ref{eq:linVlasov}), the term proportional to $\boldsymbol{v}\times\B_{0}\cdot\nabla_{v}\F$
vanishes because of the gyrotropic background distribution. The general
procedure now is to insert the plane wave expansion into the above
equations, and then derive a system of equations for $\n,$ $\dby$
and $\dbp$, using Eqs.~(\ref{eq:linE}) and (\ref{eq:linFar}) to
eliminate any dependencies on the other perturbed quantities. In the
following, we will use only the Fourier coefficients, omitting the
$k$ index and the tildes denoting the perturbed quantities. 

As a first step, we may apply the Fourier expansion to Eq.~(\ref{eq:linE})
and insert the result into Faraday's law, Eq.~(\ref{eq:linFar}).
This allows us to eliminate the electric field $\de$ and, after some
algebra, to solve for 
\begin{eqnarray}
\fl\u_{\i}\t\B_{0}=\left(\frac{\im\eta c^{2}}{4\pi}+\frac{\omega}{k^{2}}\right)\k\t\db+\frac{\im c}{4\pi en_{0\i}}\kp B_{0}\db\label{eq:uxB}\\
-\k\left[\left(\frac{\im\eta c^{2}}{4\pi}+\frac{\omega}{k^{2}}\right)\frac{\kx}{\kp}\delta B_{y}+\frac{\im c}{4\pi en_{0\i}}B_{0}\delta B_{\|}\right]\nonumber 
\end{eqnarray}
Next, we expand the Vlasov equation, Eq.~(\ref{eq:linVlasov}), in
plane waves, yielding 
\begin{equation}
\fl\im\left(\vv\d\k-\omega\right)\f=-\left[\frac{e}{\m}\left(\de+\frac{\vv\times\db}{c}\right)\right]\cdot\gv\F-\frac{e}{\m}\left(\frac{\vv\times\B_{0}}{c}\right)\cdot\gv\f.\label{eq:Vlasov_lin}
\end{equation}
Note that the term proportional to $\vv\times\db\d\gv\F$ in Eq.~(\ref{eq:Vlasov_lin})
vanishes only in an isotropic Maxwellian plasma, and thus must be
kept for the present purposes. Introducing cylindric coordinates $\left(\vp\cos\theta,\vp\sin\theta,\vpa\right)$
in velocity space, we can obtain the relations
\[
\frac{e}{\m c}\vv\times\B_{0}\cdot\gv\f=-\oci\frac{\partial\f}{\partial\theta}
\]
and

\[
\fl\left[\frac{e}{\m}\frac{\vv\times\db}{c}\right]\cdot\gv\F=\frac{e}{c}\vpa F_{0}\frac{\tpa-\tp}{\tpa\tp}\left(v_{\perp}\sin\theta\boldsymbol{e}_{x}-v_{\perp}\cos\theta\boldsymbol{e_{y}}\right)\cdot\db.
\]
Inserting these, the linearized Ohm's law from Eq.~(\ref{eq:linE}),
and Eq.~(\ref{eq:uxB}) into Eq.~(\ref{eq:Vlasov_lin}), we may
write

\begin{eqnarray}
\fl\im\left(\omega-\vv\cdot\k\right)\f+\oci\frac{\partial\f}{\partial\theta}=\nonumber \\
\Biggl\{\left[\frac{1}{c}\vpa v_{\perp}\frac{\tpa-\tp}{\tpa\tp}-\left(\frac{\im\eta c}{4\pi}\frac{\kx^{2}}{\kp}+\frac{\omega}{c\kp}\right)\frac{v_{\perp}}{\tp}\right]\frac{\delta B_{y}}{B_{0}}\cos\theta\nonumber \\
+\left(\frac{1}{c}\vpa v_{\perp}\frac{\kp}{\kx}\frac{\tpa-\tp}{\tpa\tp}-\frac{\omega}{c\kx}\frac{v_{\perp}}{\tp}\right)\frac{\delta B_{\|}}{B_{0}}\sin\theta\nonumber \\
+\left[\frac{T_{0\e}}{eB_{0}}\gamma\im\left(\frac{\kx v_{\perp}}{\tp}\cos\theta+\frac{\kp\vpa}{\tpa}\right)\frac{\n}{n_{0\i}}-\frac{\im\eta c}{4\pi}\frac{\kx\vpa}{\tpa}\frac{\delta B_{y}}{B_{0}}\right]\Biggr\} eB_{0}\F\nonumber \\
\label{eq:Vlasov-spelled-out}
\end{eqnarray}
Now, we define a function $g\left(\theta\right)=\mathrm{RHS}\left[\mathrm{Eq.\ \left(\ref{eq:Vlasov-spelled-out}\right)}\right]/\Omega_{c}$,
so that this equation reduces to the form

\begin{equation}
p(\theta)\f+\frac{\partial\f}{\partial\theta}=g\left(\theta\right).\label{eq:intefac}
\end{equation}
where we set 
\[
p\left(\theta\right)=\frac{\im}{\oci}\left(\omega-\vv\d\k\right)=\frac{\im}{\oci}\left(\omega-\kp\vpa-\kx v_{\perp}\cos\theta\right).
\]
We may now obtain a solution for Eq.~(\ref{eq:intefac}) by means
of the integrating factor method, as described similarly also in Ref.~\cite{Kazeminezhad92}.
The integrating factor is defined as

\[
\mu\left(\theta\right)=\exp\left(\int p\left(\theta\right)d\theta\right)=\exp\left(\im\left(\nu\theta-\kappa\sin\theta\right)\right),
\]
with $\nu=\left(\omega-\kp\vpa\right)/\oci$ and $\kappa=\kx\vp/\oci$.
The solution for the distribution function is then given by
\begin{equation}
\f=\frac{1}{\mu\left(\theta\right)}\int_{-\sigma\oo}^{\theta}\mu\left(\theta'\right)g\left(\theta'\right)d\theta'=\int_{-\sigma\oo}^{\theta}\mathcal{K}\left(\theta,\theta'\right)g\left(\theta'\right)d\theta'.\label{eq:solution_df1}
\end{equation}
Here, we introduced the abbreviation
\[
\mathcal{K}\left(\theta,\theta'\right)=\exp\left(\im\nu\left(\theta'-\theta\right)+\im\kappa\left(\sin\theta-\sin\theta'\right)\right)
\]
to make the following equations more concise. Note that we chose the
lower integration boundary as $-\sigma\infty$, where we set $\sigma=1$
when $\Im\left(\omega\right)>0$ and $\sigma=-1$ when $\Im\left(\omega\right)<0$.
This method ensures that the integration does not contain diverging
parts, but at the same time precludes convergence of the integral
for marginally stable modes. In the numerical solution of the dispersion
relation, we will thus take steps to ensure that the solver cannot
move arbitrarily close to zero (see Sec.~\ref{sec:Hydros}). 

Using the above expression for the ion distribution function, we may
write the perturbed density as the velocity space integral of Eq.~(\ref{eq:solution_df1}),
\begin{equation}
\n=\int_{\mathcal{V}}\int_{-\sigma\oo}^{\theta}\mathcal{K}\left(\theta,\theta'\right)g\left(\theta'\right)d\theta'd^{3}v,\label{eq:dens-1}
\end{equation}
which constitutes the first equation of the matrix system we will
use to determine the wave solutions of the hybrid-kinetic system.
Two more equations may be obtained from the $x$ and $y$ component
of Eq.~(\ref{eq:uxB}). The former yields

\begin{eqnarray}
\fl n_{0\i}u_{y}=\int_{\mathcal{V}}\int_{-\oo}^{\theta}\mathcal{K}\left(\theta,\theta'\right)v_{\perp}\sin\theta g\left(\theta'\right)d\theta'd^{3}v=-n_{0\i}\left(\frac{\im\eta c^{2}}{4\pi}+\frac{\omega}{k^{2}}\right)\kp\frac{\delta B_{y}}{B_{0}}\label{eq:uy}\\
-\frac{\im c}{4\pi e}\frac{\kp^{2}}{\kx}B_{0}\frac{\delta B_{\|}}{B_{0}}-\kx\left[n_{0\i}\left(\frac{\im\eta c^{2}}{4\pi}+\frac{\omega}{k^{2}}\right)\frac{\kx}{\kp}\frac{\delta B_{y}}{B_{0}}+\frac{\im c}{4\pi e}B_{0}\frac{\delta B_{\|}}{B_{0}}\right],\nonumber 
\end{eqnarray}
and from the latter we can calculate 
\begin{eqnarray}
\fl-n_{0\i}u_{x}=-\int_{\mathcal{V}}\int_{-\oo}^{\theta}\mathcal{K}\left(\theta,\theta'\right)v_{\perp}\cos\theta g\left(\theta'\right)d\theta'd^{3}v=\frac{\im c}{4\pi e}\kp B_{0}\frac{\delta B_{y}}{B_{0}}\nonumber \\
-n_{0\i}\left(\frac{\im\eta c^{2}}{4\pi}+\frac{\omega}{k^{2}}\right)\left(\kp^{2}\frac{\delta B_{\|}}{B_{0}}/\kx+\kx\frac{\delta B_{\|}}{B_{0}}\right).\label{eq:ux}
\end{eqnarray}
In order to obtain a solution for this system of equations, we need
to solve a set of integrals
\begin{equation}
I_{abc}=\int_{\mathcal{V}}\int_{-\sigma\oo}^{\theta}\mathcal{K}\left(\theta,\theta'\right)\vpa^{a}\F f(\theta)h(\theta')d\theta'd^{3}v,\label{eq:integrals}
\end{equation}
where $a\in\left\{ 0,1\right\} $. We choose $b\in\left\{ 0,1,2\right\} $
to denote the three choices for $f\left(\theta\right)\in\left\{ 1,\vp\cos\theta,\vp\sin\theta\right\} $,
and $c\in\left\{ 0,1,2\right\} $ to denote the three choices for
$h\left(\theta'\right)\in\left\{ 1,\vp\cos\theta',\vp\sin\theta'\right\} $.
Finally, the velocity space volume element is given by $d^{3}v=\vp d\vpa d\vp d\theta$,
and $\mathcal{V}$ is used to denote the velocity space itself. Our
present system of equations requires 15 of the 18 combinations defined
thus. The solutions of these integrals can be obtained in the same
fashion as described in Ref.~\cite{Kazeminezhad92}, and is detailed
in Appendix~\ref{sec:Integrals}. Once the integrals are solved,
we can express Eqs.~(\ref{eq:dens-1})--(\ref{eq:ux}) as a matrix
equation of the form $$\left(\matrix{M&N&O\cr P&Q&R\cr S&T&U}\right)\left(\matrix{\delta B_y/B_0\cr\delta B_\|/B_0\cr\delta n_i/n_0}\right)=0.$$Introducing
the abbreviations 
\begin{eqnarray*}
A & = & -\frac{\m}{\tp}\left(\frac{\im\eta c^{2}}{4\pi}\frac{\kx^{2}}{\kp}\right)\\
B & = & -\frac{\m}{\tp}\frac{\omega}{\kp}\\
C & = & \frac{\m}{\tp}\left(\frac{\tpa-\tp}{\tpa}\right)\\
D & = & \frac{1}{\tp}\frac{T_{0\e}}{\oci}\gamma\im\kx\\
E & = & n_{0\i}\left(\frac{\im\eta c^{2}}{4\pi}+\frac{\omega}{k^{2}}\right)\\
F & = & \frac{\im c}{4\pi e}B_{0}
\end{eqnarray*}
we may write $g\left(\theta\right)$ as 
\begin{eqnarray*}
\fl g\left(\theta\right)=\left\{ \frac{\delta B_{y}}{B_{0}}\left[A+B+C\vpa\right]+\frac{\n}{n_{0\i}}D\right\} \vp\F\cos\theta\\
+\frac{\delta B_{\|}}{B_{0}}\frac{\kp}{\kx}\left[B+C\vpa\right]\vp\F\sin\theta+\frac{\kp}{\kx}\frac{\tp}{\tpa}\left[\frac{\delta B_{y}}{B_{0}}A+\frac{\n}{n_{0\i}}D\right]\vpa\F,
\end{eqnarray*}
and simplifying also Eqs.~(\ref{eq:dens-1})--(\ref{eq:ux}), we
finally obtain the matrix elements 
\[
M=E\frac{k^{2}}{\kp}+\left(A+B\right)I_{021}+CI_{121}+AI_{120}\frac{\kp}{\kx}\frac{\tp}{\tpa}
\]
\[
N=F\frac{k^{2}}{\kx}+BI_{022}\frac{\kp}{\kx}+CI_{122}\frac{\kp}{\kx}
\]
\[
O=D\left(I_{120}\frac{\kp}{\kx}\frac{\tp}{\tpa}+I_{021}\right)
\]
\[
P=F\kp+\left(A+B\right)I_{011}+CI_{111}+AI_{110}\frac{\kp}{\kx}\frac{\tp}{\tpa}
\]
\[
Q=-E\frac{k^{2}}{\kx}+BI_{012}\frac{\kp}{\kx}+CI_{112}\frac{\kp}{\kx}
\]
,
\[
R=D\left(I_{110}\frac{\kp}{\kx}\frac{\tp}{\tpa}+I_{011}\right)
\]
\[
S=\left(A+B\right)I_{001}+CI_{101}+AI_{100}\frac{\kp}{\kx}\frac{\tp}{\tpa}
\]
\[
T=BI_{002}\frac{\kp}{\kx}+CI_{102}\frac{\kp}{\kx}
\]
\[
U=-n_{0\i}+D\left(I_{100}\frac{\kp}{\kx}\frac{\tp}{\tpa}+I_{001}\right)
\]
The dispersion relation of the hybrid-kinetic model is then obtained
by setting \begin{equation}\det\left(\matrix{M&N&O\cr P&Q&R\cr S&T&U}\right)=0,\label{eq:DR}\end{equation}
and solving for the complex frequencies that fulfill this equation.

\section{HYDROS: a \uline{Hy}brid \uline{D}ispersion \uline{R}elati\uline{o}n
\uline{S}olver}

\label{sec:Hydros}

\subsection{Numerical implementation}

The dispersion relation derived in Sec.~\ref{sec:deriv} has been
implemented for numerical solution using the Python/NumPy/SciPy framework,
whose mathematical library provides all of the necessary functionality.
Both the plasma dispersion function and the modified (and exponentially
scaled) Bessel functions which appear when solving the integrals of
Eq.~(\ref{eq:integrals}) are provided by SciPy \cite{Jones01} via
the specfun library \cite{Cody93}, and can thus be readily used.
In order to find the zeros of the dispersion relation, we employ the
root finding methods provided by the SciPy library, with a Levenberg-Marquardt
algorithm \cite{Levenberg44,Marquardt63} set as the default method. 

To improve the convergence speed of the root finding algorithm, several
predictive algorithms are available to anticipate the evolution of
a root during a parameter scan, or to direct the solver closer towards
a particular (e.g. less damped) solution when several roots exist
close together. These algorithms are a) using the old position of
the root as a starting point for the new iteration, b) quadratic extrapolation
using the last two positions of the root, c) modified quadratic extrapolation.
The last method consists of a quadratic extrapolation with a subsequent
modification applied to the predicted frequency or damping rate and
can, e.g., be used for cases with degenerate modes. For all cases
shown here, the dispersion relations were sufficiently clear that
method b) could be employed without additional modifications. However,
we found it very useful to incorporate diagnostics to ensure that
the solver keeps following a particular solution, e.g., by comparing
complex frequencies, field amplitudes, and cross phases between the
fields. 

As mentioned in Sec.~\ref{sec:deriv}, the integrals solved here
do not converge if the imaginary part of the frequency is zero. For
that reason, we introduce a lower boundary $\gamma_{\mathrm{min}}=10^{-13}$
for the absolute value of the imaginary part. If the solver converges
to a number smaller than $\gamma_{\mathrm{min}}$, the starting point
for the next solution will be reset to $\gamma_{\mathrm{min}}$. This
prevents a runaway of the solver towards ever smaller imaginary parts
that has been observed otherwise.

\subsection{Benchmark of HYDROS against other kinetic solvers}

\label{sec:benchmark}

In this section, we aim to verify both the derivation of Sec.~\ref{sec:deriv}
and the numerical implementation of the HYDROS code by comparing its
results to solutions of the fully kinetic dispersion solver DSHARK
\cite{Astfalk15}. For this exercise, we choose several example waves
and instabilities that are commonly encountered in many systems of
interest (e.g., the solar wind and planetary magnetospheres), such
as the fast and slow magnetosonic mode, the kinetic Alfvén wave (KAW),
the oblique and parallel firehose mode and the mirror mode. For all
studies in this work, we set the polytropic coefficient $\gamma=1$
to obtain an isothermal electron description, and we choose the resistivity
$\eta=0$. This section will be subdivided by the propagation angle
of the examined modes. We note that neither DSHARK nor HYDROS are
able to treat exact parallel and perpendicular propagation, as these
cases lead to divisions by zero in the general expressions for the
dispersion relation. ``Parallel'' and ``perpendicular'' thus refer
to an angle very close to 0 degrees or 90 degrees, respectively. In
DSHARK, we furthermore set $\va/c=0.01$ throughout (this ratio is
undefined in HYDROS, as the displacement current is neglected).

In this section, wavenumbers will be plotted in units of $kd_{\i}$,
where $d_{\i}=c/\omega_{\mathrm{pi}}=\sqrt{m_{\i}c^{2}/4\pi n_{\i}e^{2}}$,
and frequencies will be given in terms of the ion cyclotron frequency
$\oci=eB_{0}/m_{\i}c$. The plasma beta parameter for a species $s$
will be defined as $\beta_{s}=8\pi n_{s}T_{s}/B_{0}^{2}.$

\subsubsection{Parallel propagation}

\subsubsection*{Parallel firehose instability.}

\begin{figure}
\subfloat[\label{fig:PFI}]{\includegraphics[width=0.5\textwidth]{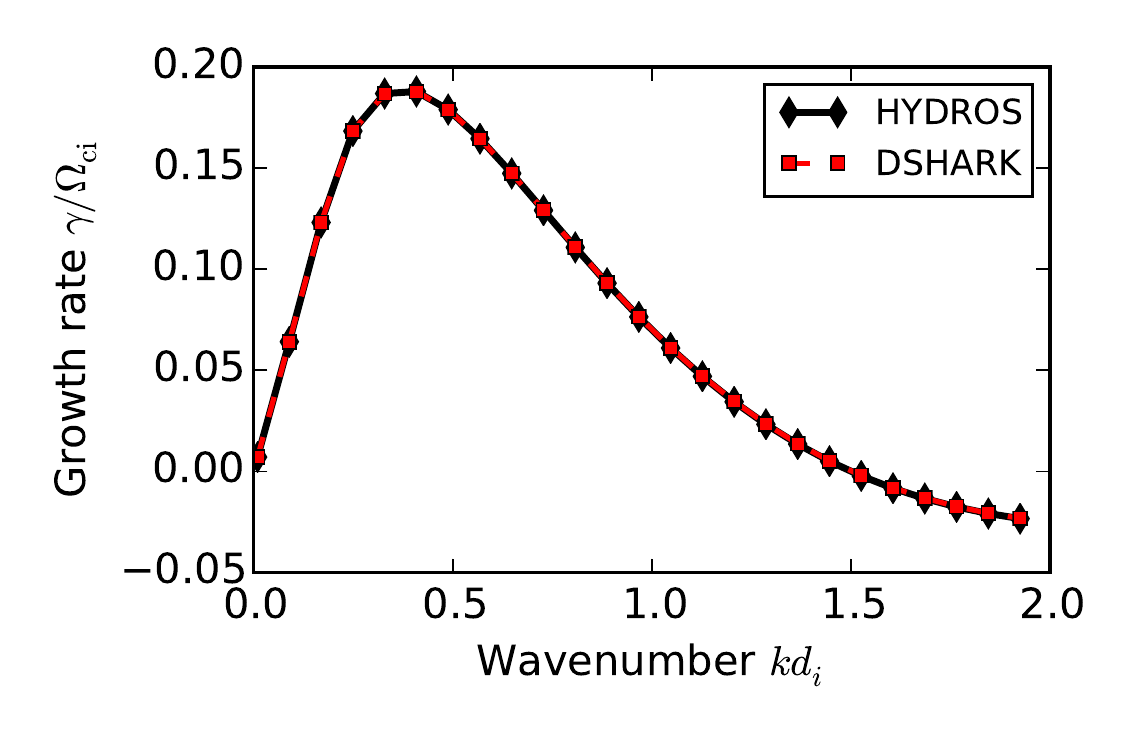}

}\subfloat[\label{fig:PFI-1}]{\includegraphics[width=0.5\textwidth]{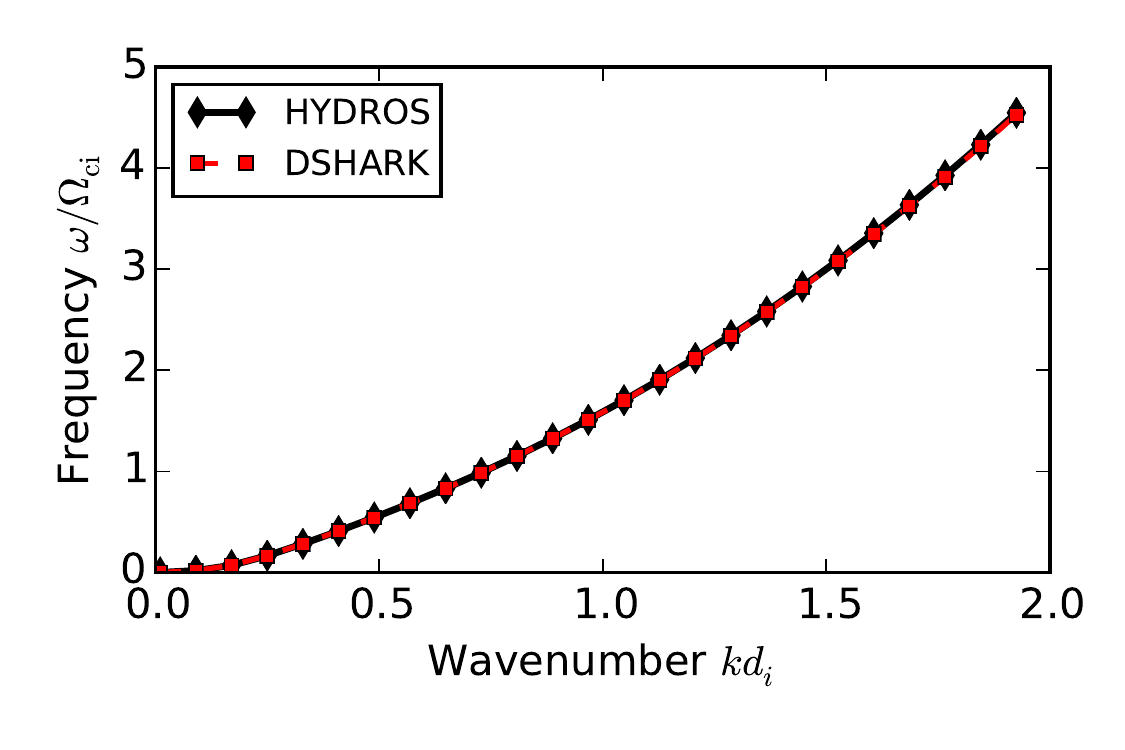}

}\caption{Growth rates (a) and frequencies (b) for parallel firehose instability
solutions of HYDROS and DSHARK, using $\beta_{\protect\p\|}=4$ and
$\beta_{\protect\p\perp}=\beta_{\protect\e\|}=1$. }

\end{figure}

As the first example, we compare growth rates and frequencies of the
parallel firehose instability, which occurs when the parallel thermal
pressure significantly exceeds the perpendicular thermal pressure.
Here, we choose the parameters $\beta_{\p\|}=4$, $\beta_{\p\perp}=\beta_{\e\|}=1$,
and we set the propagation angle to $10^{-4}$ degrees. The results
for these parameters from both the fully kinetic and the hybrid-kinetic
solver are shown in Figs.~\ref{fig:PFI} and \ref{fig:PFI-1}. Both
in growth rates and frequencies we find very good agreement between
the two solvers.

\subsubsection*{Ion cyclotron instability.}

\begin{figure}
\subfloat[\label{fig:ICI}]{\includegraphics[width=0.5\textwidth]{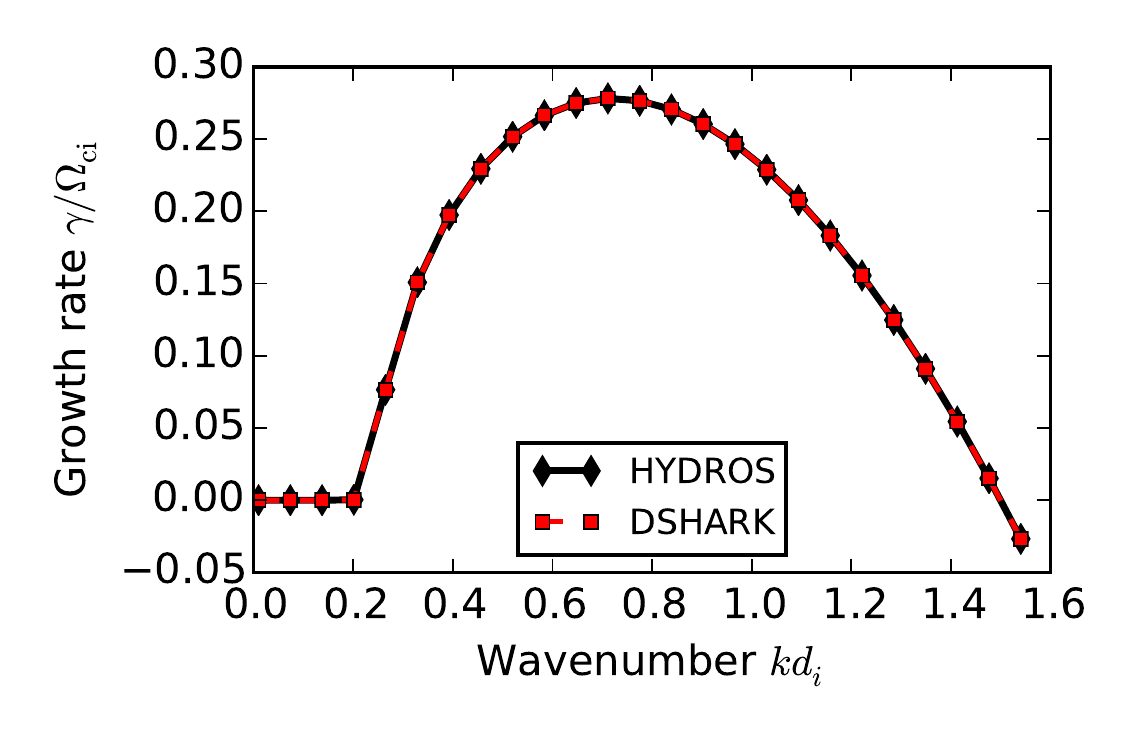}

}\subfloat[\label{fig:ICI-1}]{\includegraphics[width=0.5\textwidth]{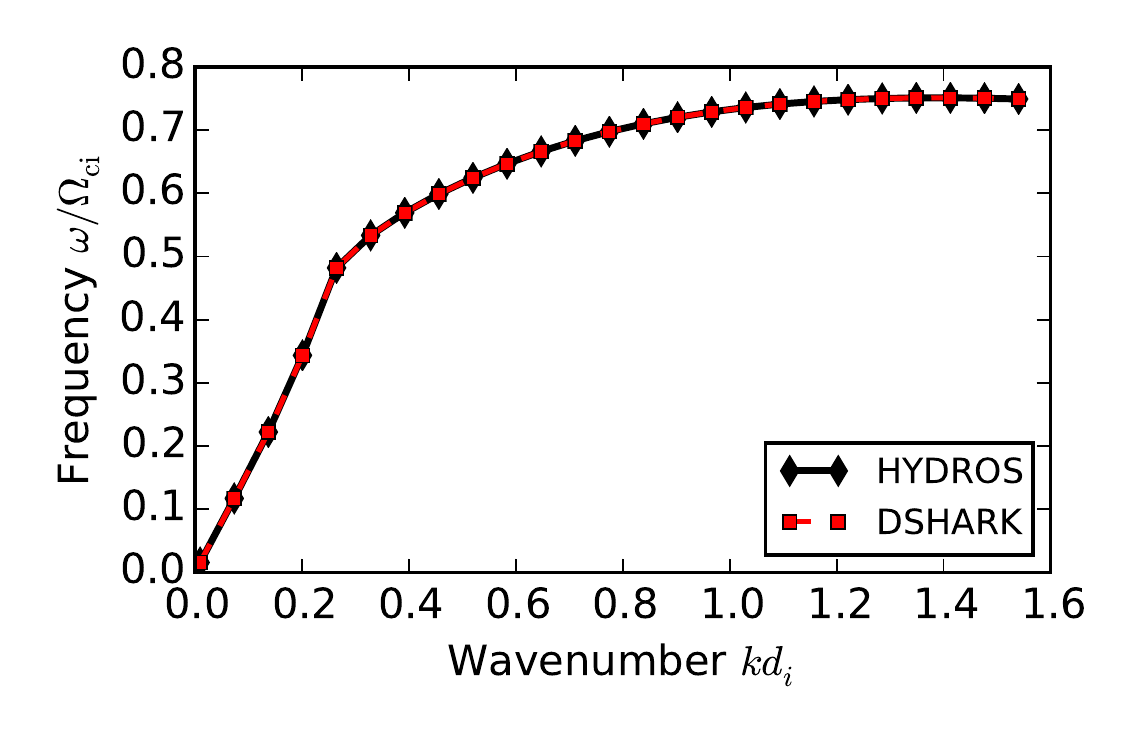}

}

\caption{Growth rates (a) and frequencies (b) for ion cyclotron instability
solutions of HYDROS and DSHARK, using $\beta_{\protect\p\|}=\beta_{\protect\e}=1$
and $\beta_{\protect\p\perp}=4$.}
\end{figure}

Inverting the anisotropy used for the parallel firehose instability
to $\beta_{\p\|}=\beta_{\e}=1$, $\beta_{\p\perp}=4$ while keeping
the propagation angle parallel to the background field yields a plasma
that is susceptible to the ion cyclotron instability. Scanning over
the wavenumber (see Figs.~\ref{fig:ICI} and \ref{fig:ICI-1}) then
reveals a growing wave with a frequency close to the ion cyclotron
frequency, and a growth rate peak close to $kd_{\i}=0.7$. As before,
DSHARK and HYDROS agree very well on both growth rates and frequencies.

\subsubsection*{L-mode.}

\begin{figure}
\subfloat[\label{fig:Ldamp}]{\includegraphics[width=0.5\textwidth]{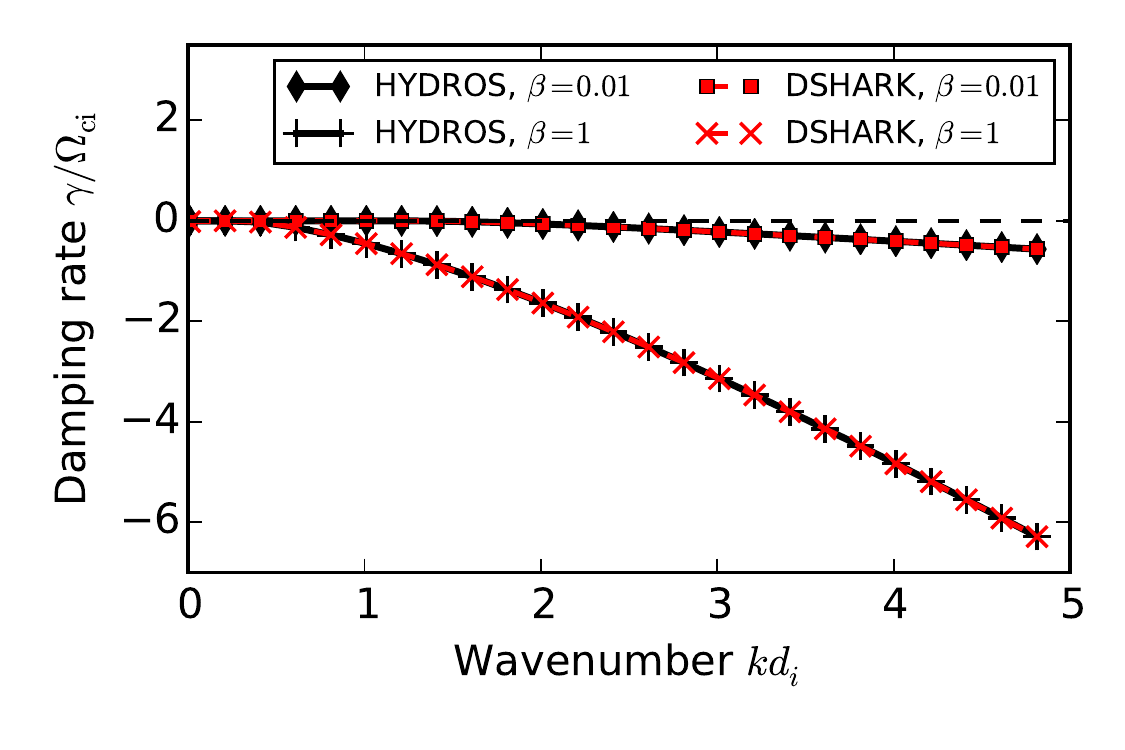}

}\subfloat[\label{fig:Lfreq}]{\includegraphics[width=0.5\textwidth]{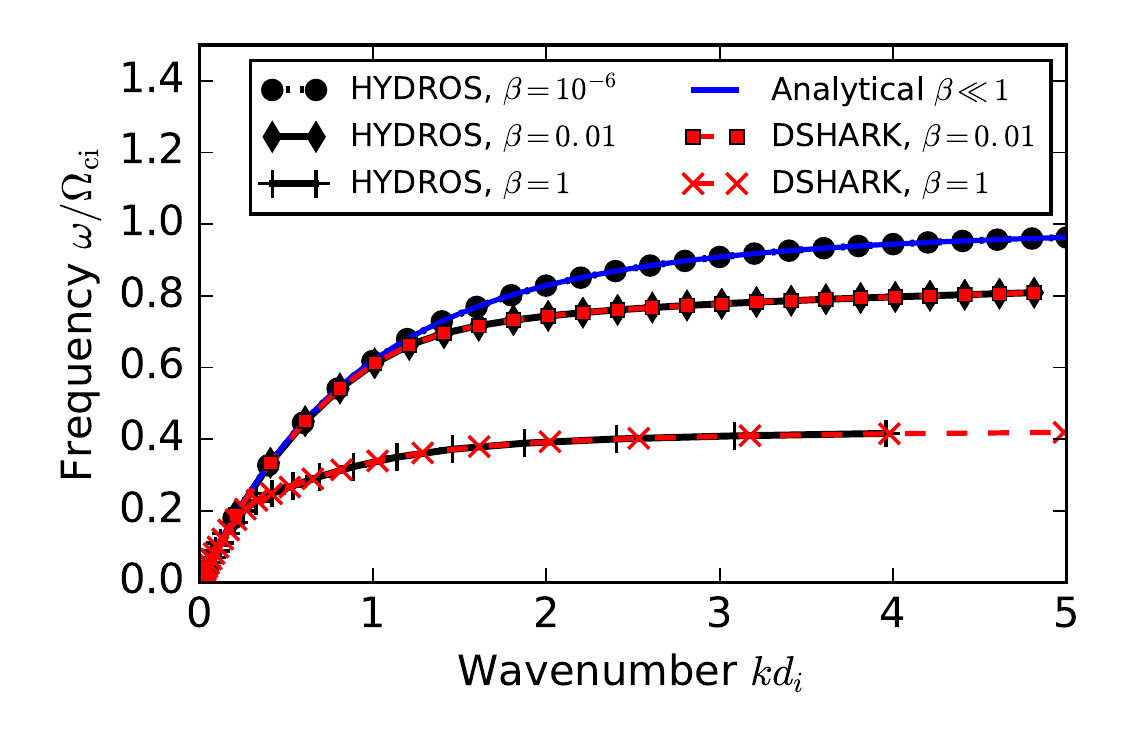}

}

\caption{L-mode damping rates (a) and frequencies (b) for $\beta=10^{-6},\ 0.01,\ 1.0$,
for fully kinetic and hybrid-kinetic descriptions. For comparison,
an analytical frequency formula for low $\beta$ \cite{Valentini07}
is shown in blue. }

\end{figure}

Next, we choose an isotropic ion distribution and examine the results
for the L-mode solution ($\Alf$/ion-cyclotron wave). Here, we select
two different values of the plasma $\beta$, namely $\beta_{\i}=\beta_{\e}=10^{-2}$
and $\beta_{\i}=\beta_{\e}=1$. The relevance of the $\beta$ value
in this case is its influence on the effect of the ion cyclotron resonance
-- for low $\beta$, one indeed finds that the wave frequency asymptotes
close to $\oci$, whereas it converges towards a lower frequency in
the finite-$\beta$ case, accompanied by a significantly stronger
damping. Once again, both DSHARK and HYDROS agree very well on both
this effect and on the dispersion curves for the two cases, which
are plotted in Figs.~\ref{fig:Ldamp} and \ref{fig:Lfreq}. Furthermore,
for low $\beta$, very good agreement with an analytical fluid dispersion
relation \cite{Valentini07} is found. 

\subsubsection*{R-mode.}

\begin{figure}
\subfloat[\label{fig:Rdamp}]{\includegraphics[width=0.5\textwidth]{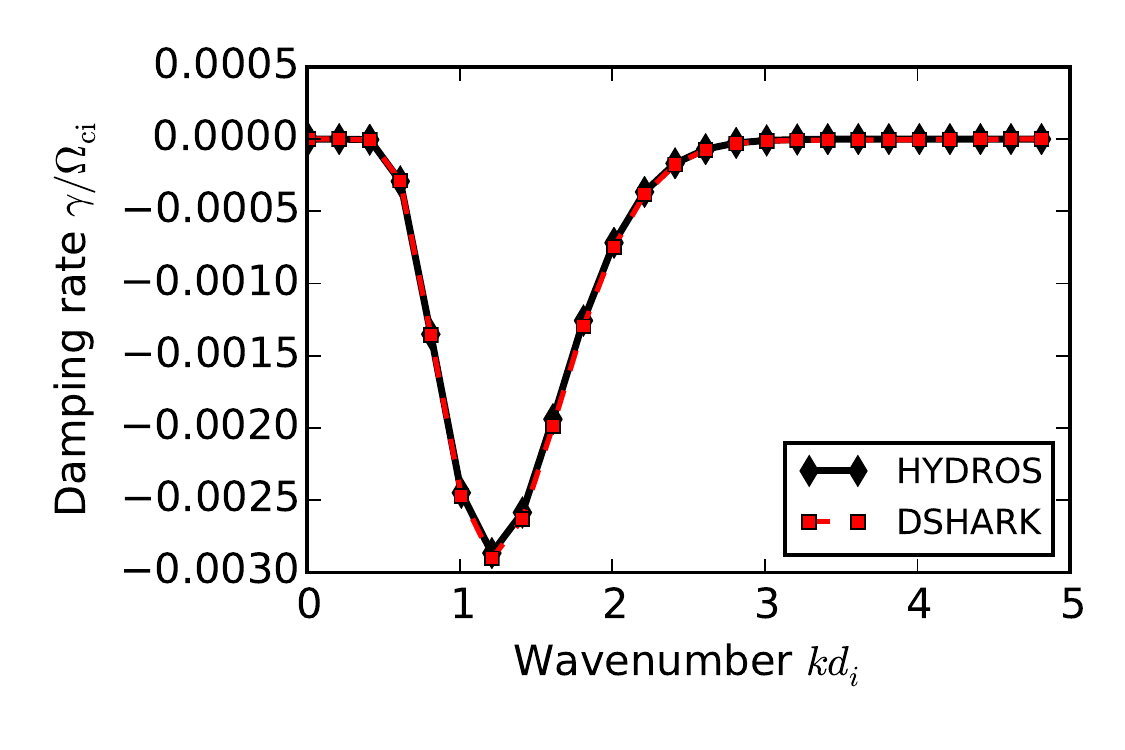}

}\subfloat[\label{fig:Rfrq}]{\includegraphics[width=0.5\textwidth]{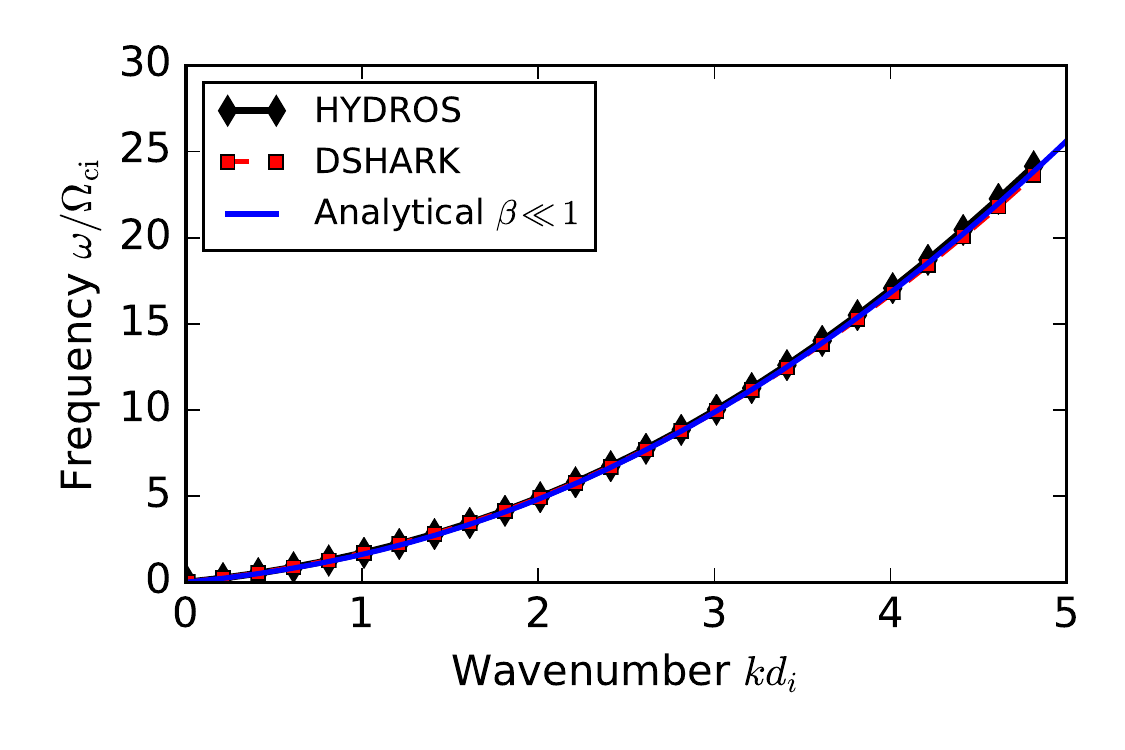}

}

\caption{R-mode damping rates (a) and frequencies (b) for HYDROS and DSHARK,
using $\beta_{\protect\i\|}=\beta_{i\perp}=\beta_{\protect\e}=1$.
For comparison, an analytical R-mode frequency formula \cite{Valentini07}
is plotted for the limit of zero electron mass and low $\beta$ (blue).}
\end{figure}

The R-mode (or whistler) is characterized by its dispersive behavior,
with frequencies (for parallel propagation) proportional to $k_{\|}\va$
for $k_{\|}\di\leq1$, but proportional to $\left(\kp\va\right)^{2}$
at higher wavenumber. In Figs.~\ref{fig:Rdamp} and \ref{fig:Rfrq}
the damping rates and frequencies for this wave are presented, for
$\beta_{\i\|}=\beta_{i\perp}=\beta_{\e}=1$, along with an analytical
dispersion relation \cite{Valentini07}. A small frequency deviation
between DSHARK and HYDROS towards higher wavenumbers can be observed,
which follows from neglecting electron inertia effects in the hybrid-kinetic
model. The latter thus shows a perfect $\kp^{2}$ scaling towards
higher wavenumbers, whereas the fully kinetic model eventually rolls
over and asymptotes towards close to $\oce$ (not shown). This effect
is also kept in the analytical fluid dispersion relation \cite{Valentini07}
plotted as well in Fig.~\ref{fig:Rfrq}. In the damping rates of
Fig.~\ref{fig:Rdamp}, a slight discrepancy can be discerned around
$kd_{\i}\approx1$, which may be attributed to the effects of electron
Landau or Barnes damping \cite{Barnes66}, which is not retained within
the hybrid-kinetic model. The electron mass may be artificially reduced
towards the limit taken in our hybrid derivation, improving the agreement
between the two curves (not shown). 

\paragraph*{Ion acoustic waves.}

\begin{figure}
\subfloat[\label{fig:IAWdamp}]{\includegraphics[width=0.5\textwidth]{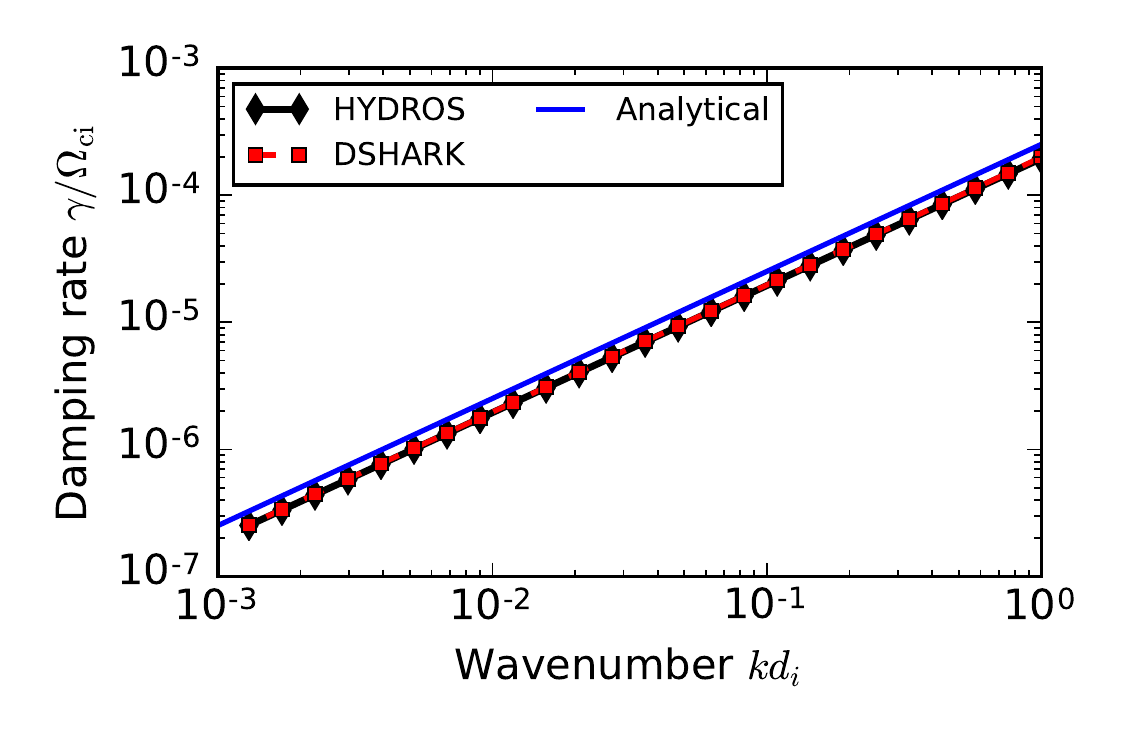}

}\subfloat[\label{fig:IAWfrq}]{\includegraphics[width=0.5\textwidth]{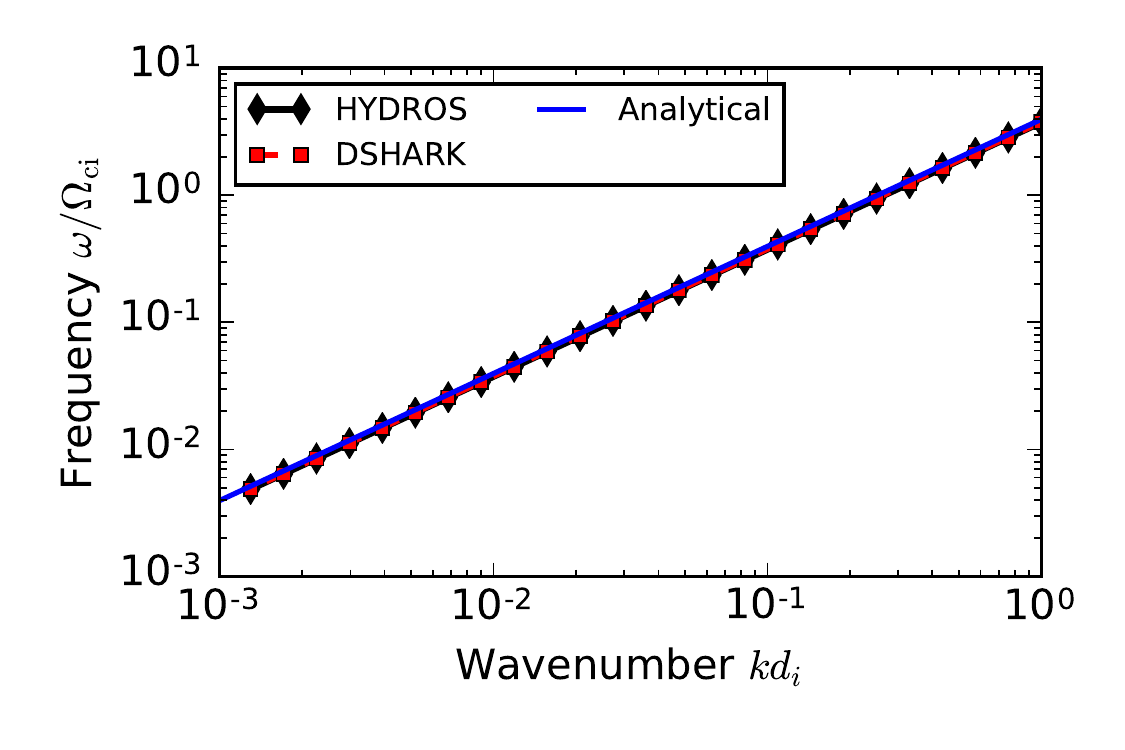}

}

\caption{Ion acoustic wave damping rates (a) and frequencies (b) for HYDROS
and DSHARK (with $m_{\protect\e}/m_{\protect\i}=10^{-12}$), using
$\beta_{\protect\i\|}=\beta_{i\perp}=1$, and $\beta_{\protect\e}=25$.
For comparison, analytical frequencies and damping rates \cite{Valentini07}
are plotted in blue, valid for $T_{\protect\i}\ll T_{\protect\e}$.}
\end{figure}

Finally, we examine the dispersion relation of the ion acoustic mode.
Although the wavevector of this mode can be arbitrarily oriented with
respect to the background field, we choose to examine it for parallel
propagation here to enable a cleaner comparison to analytical theory,
avoiding visible cyclotron damping effects. Figs.~\ref{fig:IAWdamp}
and \ref{fig:IAWfrq} present the code results, as well as analytical
formulae for damping rates and frequencies \cite{Valentini07}. Note
that the latter apply for $T_{\i}/T_{\e}\ll1$, but a relatively large
value of 0.04 has been chosen for this ratio to achieve well-measurable
damping rates. Thus, although the results from both numerical solvers
agree well with other, they deviate somewhat from the analytical damping
rates. In addition, for this wave a very small electron mass $m_{\e}=10^{-12}m_{\i}$
had to be chosen in DSHARK to avoid electron Landau damping, which
would otherwise dominate here.

\subsubsection{Perpendicular propagation}

\subsubsection*{Ion Bernstein modes.}

\begin{figure}
\subfloat[\label{fig:IBM1}]{\includegraphics[width=0.5\textwidth]{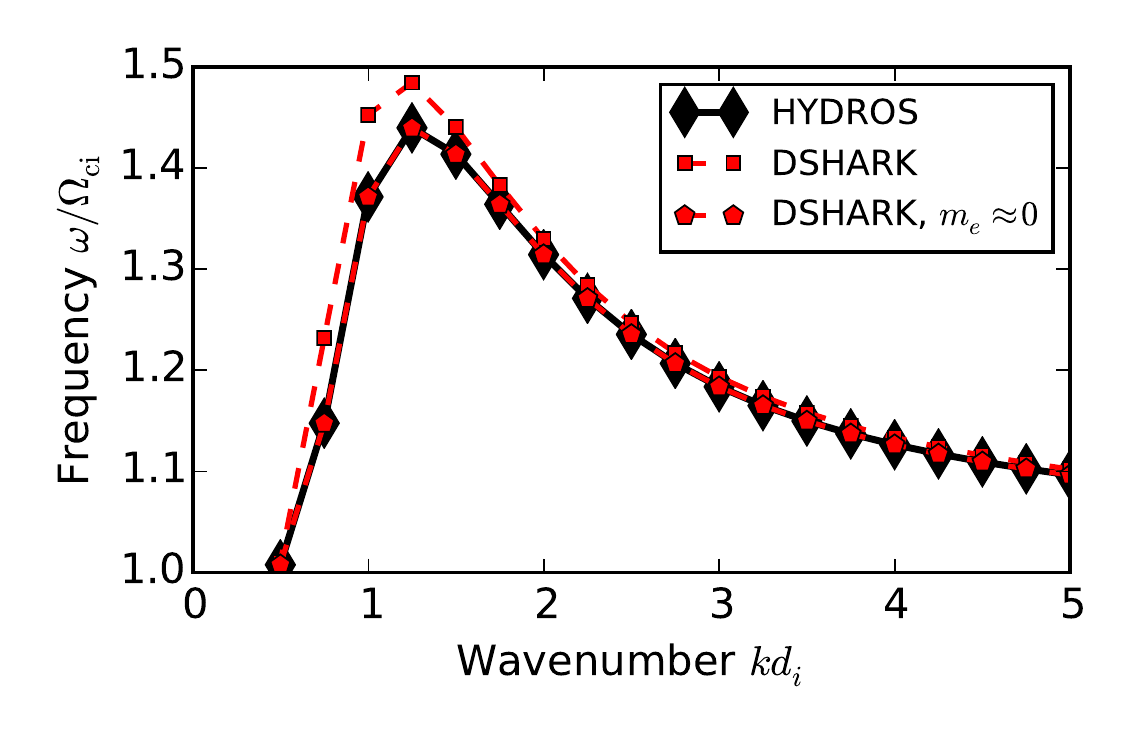}

}\subfloat[\label{fig:IBM2}]{\includegraphics[width=0.5\textwidth]{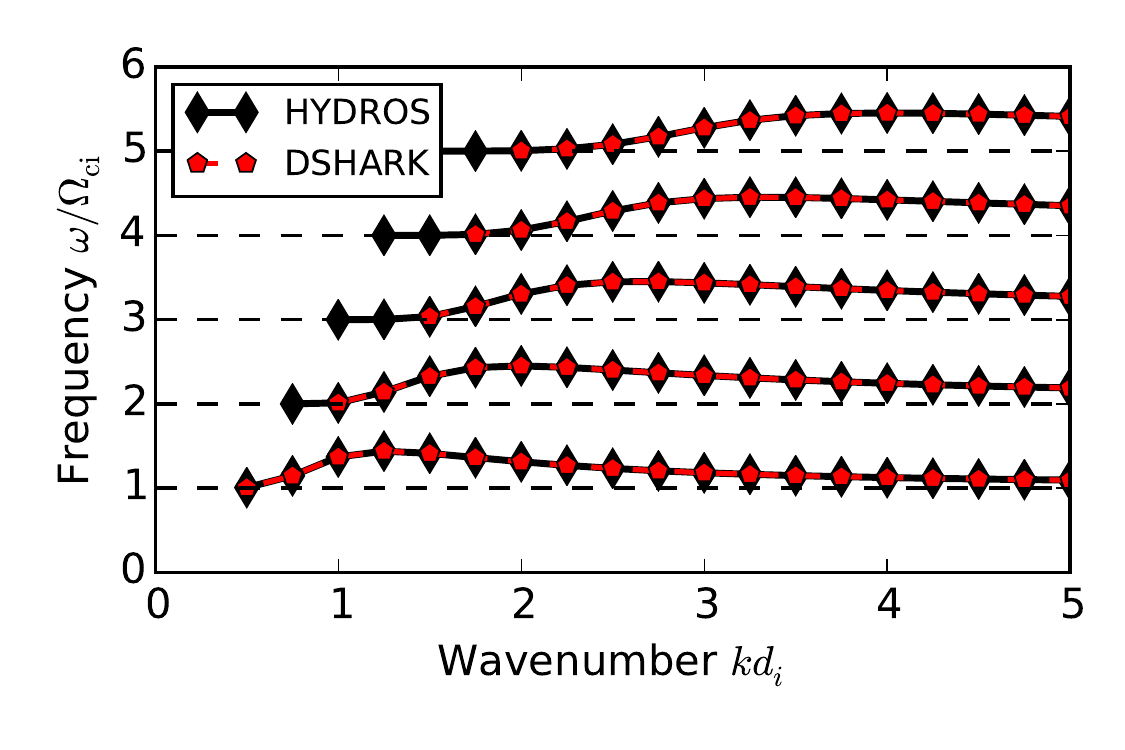}

}

\caption{(a) Ion Bernstein mode comparison to DSHARK with the actual proton/electron
mass ratio (squares), and with $m_{\protect\e}/m_{\protect\p}=10^{-15}$
(pentagons). (b) Comparison of the first five ion Bernstein modes
to DSHARK, using only $m_{\protect\e}/m_{\protect\p}=10^{-15}$. In
all cases, $\beta_{\protect\i\|}=\beta_{\protect\i\perp}=\beta_{\protect\e}=1$.}

\end{figure}

Next, we focus on purely perpendicularly propagating ion Bernstein
modes, for $\beta_{\i\|}=\beta_{\i\perp}=\beta_{\e}=1$.  For the
first ion Bernstein mode close to $\omega=\oci$, we find that its
peak frequency deviates from the DSHARK result by about $\lesssim5\%$
(see Fig.~\ref{fig:IBM1}). This discrepancy, however, is resolved
when approximating, in the kinetic dispersion solver, the massless
electron limit that enters the HYDROS equations. Good numerical agreement
is obtained for $m_{\e}/m_{\p}\lesssim10^{-12}$, and for Figs.~\ref{fig:IBM1}
and \ref{fig:IBM2} a value of $m_{\e}=10^{-15}m_{\p}$ has been used.
With this setup, excellent agreement is obtained for the five first
ion Bernstein modes, as demonstrated in Fig.~\ref{fig:IBM2}. The
damping rates obtained for these modes are numerical zeros (Landau
and transit time damping are negligible because $\kp\approx0$) and
are thus not plotted.

\subsubsection{Oblique propagation}

In this subsection, we explore the general case of oblique propagation,
in which various processes, e.g. resonant wave-particle interactions
that depend on perpendicular and parallel velocity, occur in combination,
resulting in more complex mode behavior.

\subsubsection*{Oblique firehose instability.}

\begin{figure}
\subfloat[\label{fig:OFI}]{\includegraphics[width=0.5\textwidth]{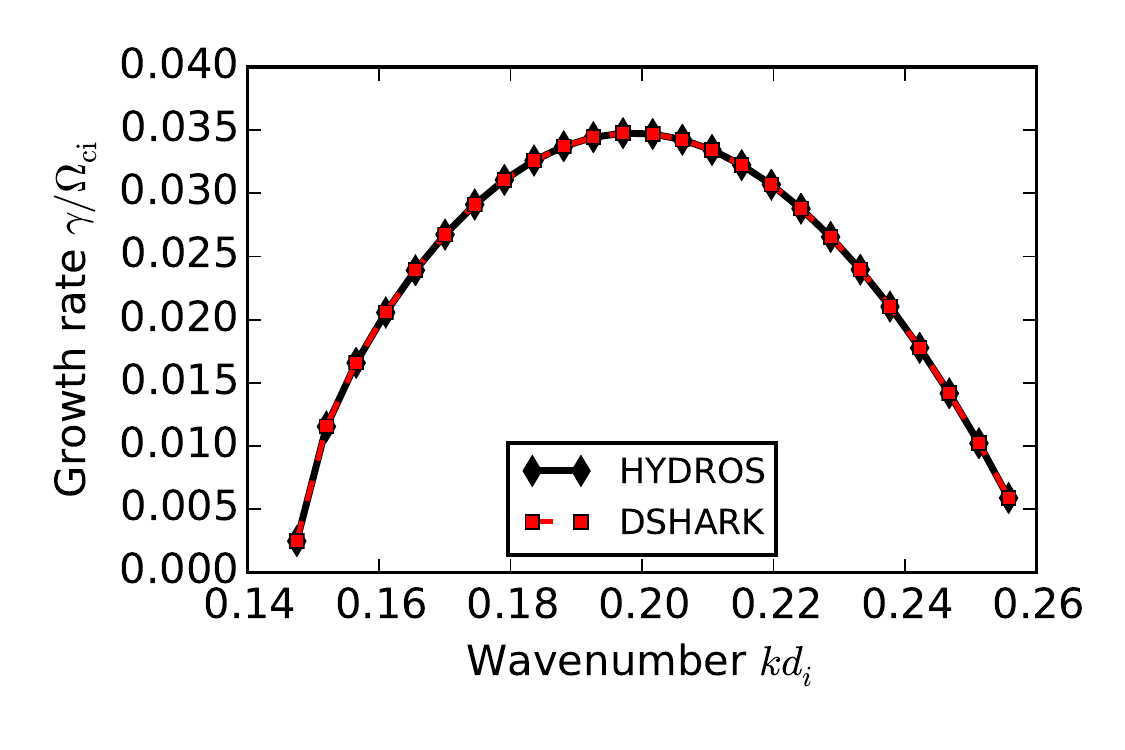}

}\subfloat[\label{fig:Mirror}]{\includegraphics[width=0.5\textwidth]{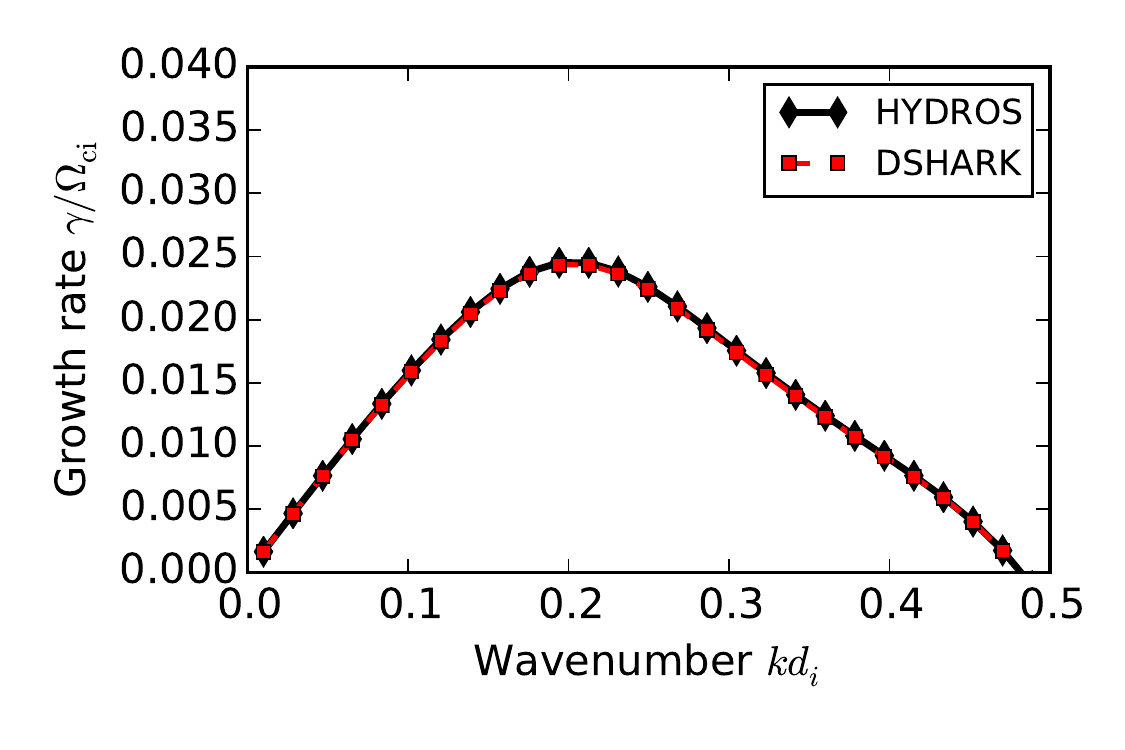}

}

\caption{Growth rate comparison for (a) the oblique firehose instability (for
$\beta_{\mathrm{i}\|}=10$, $\beta_{\mathrm{i}\perp}=26/3$, $\beta_{\protect\e}=1$),
(b) the mirror instability (using $\beta_{\protect\i\|}=6,$ $\beta_{i\perp}=10$,
$\beta_{\protect\e}=1$) between HYDROS and DSHARK. Both plots use
a propagation angle of 45$^{\circ}$.}

\end{figure}

The first instability we examine is the oblique firehose instability,
which results in a non-propagating (zero frequency) mode whose growth
rate depends on the wavevector. Here, we choose $\beta_{\mathrm{i}\|}=10$,
$\beta_{\mathrm{i}\perp}=26/3$, $\beta_{\e}=1$ and a propagation
angle of 45$^{\circ}$. In Fig.~\ref{fig:OFI}, we compare the latter
between HYDROS and DSHARK, showing excellent agreement of the growth
rates. Frequencies are not shown due to the non-propagating nature
of this mode.

\subsubsection*{Mirror mode.}

Next, we examine the mirror mode, which responds to an anisotropy
of the ion distribution that favors the perpendicular pressure. We
set $\beta_{\i\|}=6,$ $\beta_{i\perp}=10$, $\beta_{\e}=1$, keeping
the propagation angle of 45$^{\circ}$. The resulting growth rates
are shown in Fig.~\ref{fig:Mirror} and, once more, show good agreement
between the two codes. A close look, however, shows some minor discrepancy
close to the growth rate peak. This is again a result of the electron
physics missing from the hybrid model, and can be resolved when taking
the electron towards zero in the fully kinetic code.

\subsubsection*{Kinetic $\protect\Alf$ wave.}

\begin{figure}
\subfloat[\label{fig:KAWf}]{\includegraphics[width=0.5\textwidth]{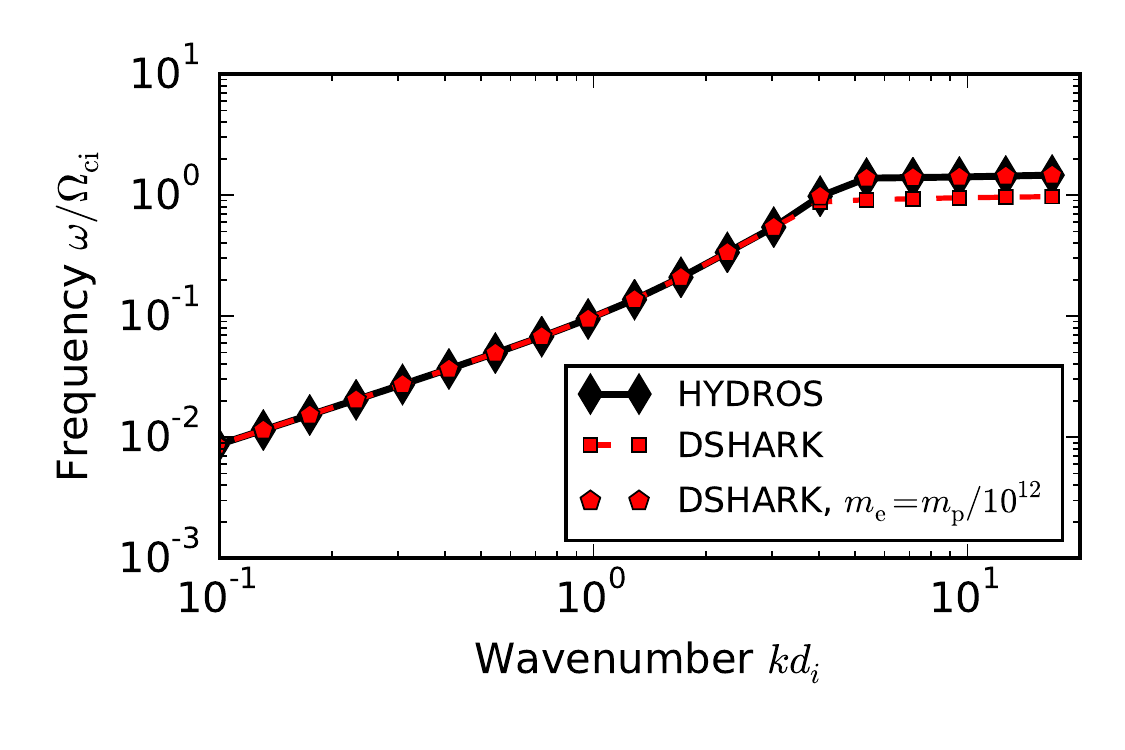}

}\subfloat[\label{fig:KAWd}]{\includegraphics[width=0.5\textwidth]{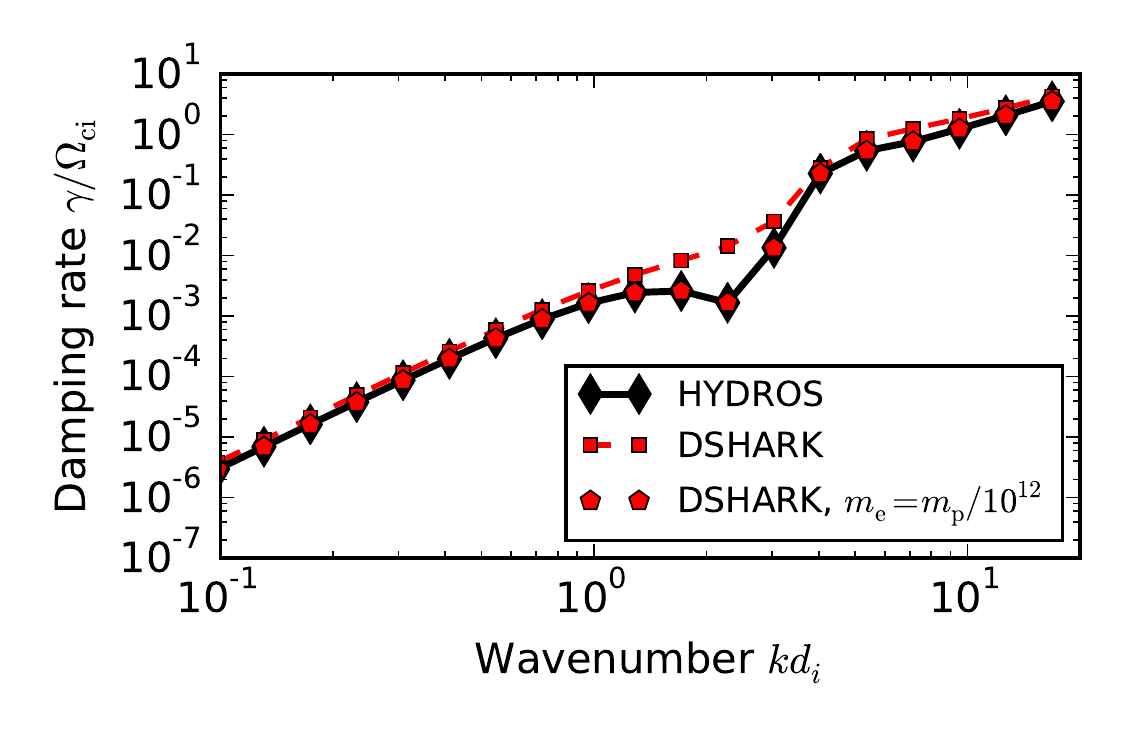}

}

\caption{Comparison of frequencies (a) and damping rates (b) for the kinetic
$\protect\Alf$ wave. DSHARK results are shown with the real proton/electron
mass ratio (squares), and with very small electron mass, $m_{\protect\e}=m_{\protect\p}/10^{12}$
(pentagons). The parameters are $\beta_{\protect\e}=\beta_{\protect\p\|}=\beta_{\protect\p\perp}=1$,
and a propagation angle of $85^{\circ}$.}
\end{figure}

The kinetic $\Alf$ wave is the kinetic equivalent of the MHD shear-$\Alf$
wave, and will be examined here for a propagation angle of $85^{\circ}$,
and $\beta_{\e}=\beta_{\p\|}=\beta_{\p\perp}=1$. We scan the wavenumbers
from $kd_{\i}=0.1$ up to $kd_{\i}=20$, using first the real proton/electron
mass ratio in DSHARK (squares in Figs.~\ref{fig:KAWf} and \ref{fig:KAWd}).
As is demonstrated in Fig.~\ref{fig:KAWf}, we find very good agreement
in the frequencies up to about $kd_{\i}\sim3$. Beyond this wavenumber,
however, the hybrid model KAW converges to a frequency of about 1.5$\oci$,
whereas in the kinetic model it stays slightly below $\oci$. 

Although barely visible in Fig.~\ref{fig:KAWd}, we find that the
hybrid model underpredicts the fully kinetic KAW damping at ion scales
(below $kd_{\i}\sim1$) by about 25\%, which can be attributed to
the missing electron Landau damping. Above $kd_{i}\sim1$, this gap
becomes even more significant, but closes again as the ion cyclotron
damping becomes dominant at about $kd_{i}\sim4$. Reducing the electron
mass in DSHARK to $m_{\e}=10^{-12}m_{\p}$, the hybrid results both
for the frequency and the damping rates are recovered with very good
accuracy (pentagons in Figs.~\ref{fig:KAWf} and \ref{fig:KAWd}). 

\subsubsection*{Fast magnetosonic mode.}

\begin{figure}
\subfloat[\label{fig:MSw}]{\includegraphics[width=0.5\textwidth]{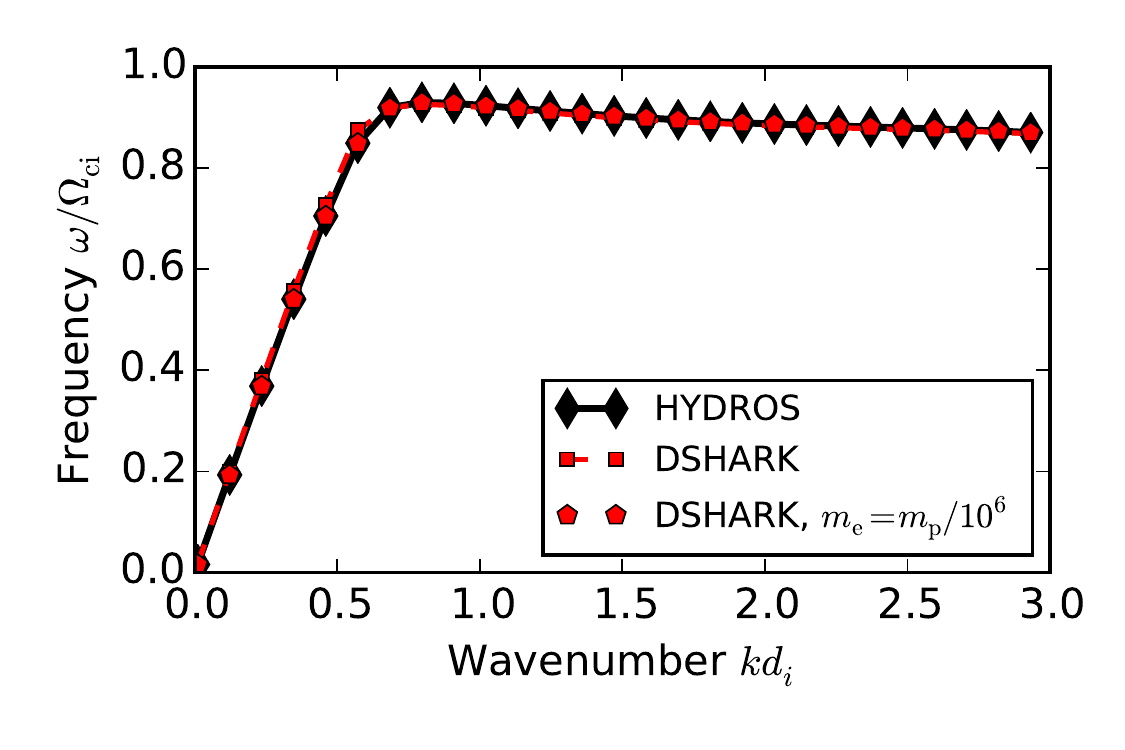}

}\subfloat[\label{fig:MSg}]{\includegraphics[width=0.5\textwidth]{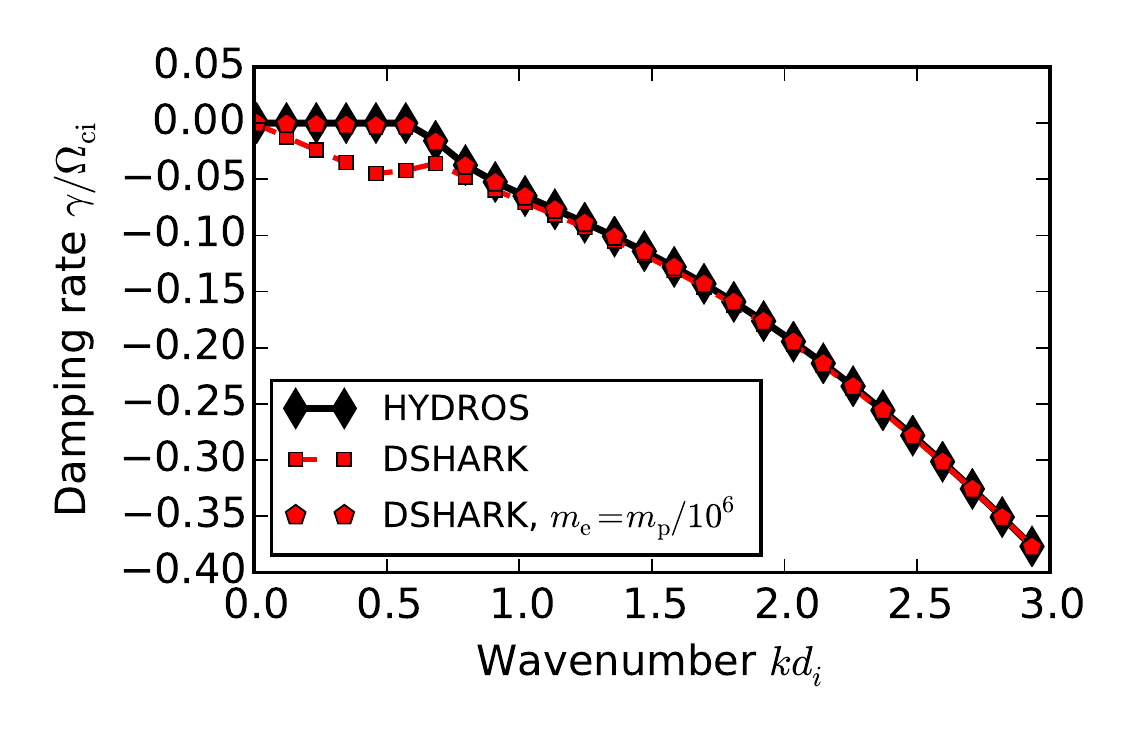}

}

\caption{Comparison of fast magnetosonic mode frequencies (a) and damping rates
(b) between HYDROS and DSHARK. DSHARK results are shown both with
the real proton/electron mass ratio (squares) and reduced electron
mass $m_{\protect\e}=m_{\protect\p}/10^{6}$ (pentagons). The parameters
are $\beta_{\protect\e}=\beta_{\protect\p\|}=\beta_{\protect\p\perp}=1$,
and a propagation angle of $85^{\circ}$.}
\end{figure}

Our next test concerns the fast magnetosonic mode, in the wavenumber
range where its frequency approaches that of the ion cyclotron motion.
For the chosen propagation angle of 85$^{\circ}$, this occurs roughly
at a wavenumber of $kd_{\i}\approx0.6$. As before, we use $\beta_{\i\|}=\beta_{\i\perp}=\beta_{\e}=1$.
For these parameters, the fast mode has a left-handed polarization,
such that it resonates with the ion gyration and is confined to frequencies
below the ion cyclotron frequency. We present the comparison of the
resulting frequencies in Fig.~\ref{fig:MSw}, and the corresponding
damping rates in Fig.~\ref{fig:MSg}. 

In the frequency plot, good agreement is observed, although at low
wavenumbers there is a slight shift between the curves of the hybrid
and the fully kinetic model until the ion cyclotron frequency is approached.
In the same wavenumber range, the damping rate plot exhibits more
severe disagreement: for proton/electron mass ratio (squares) the
fully kinetic model yields significant finite damping rates, whereas
the hybrid model predicts undamped waves. In this case, the observed
wave damping is caused by electron transit time damping \cite{Barnes66},
and proves to be rather resilient when changing the electron mass
-- in order to obtain good agreement with the hybrid model, as shown
in Fig.~\ref{fig:MSg}, the electron mass had to be reduced to $m_{\e}=m_{\p}/10^{6}$.
This modification, at the same time, removes the frequency shift observed
before.

\subsubsection*{Slow magnetosonic mode.}

\begin{figure}
\subfloat[\label{fig:SMw}]{\includegraphics[width=0.5\textwidth]{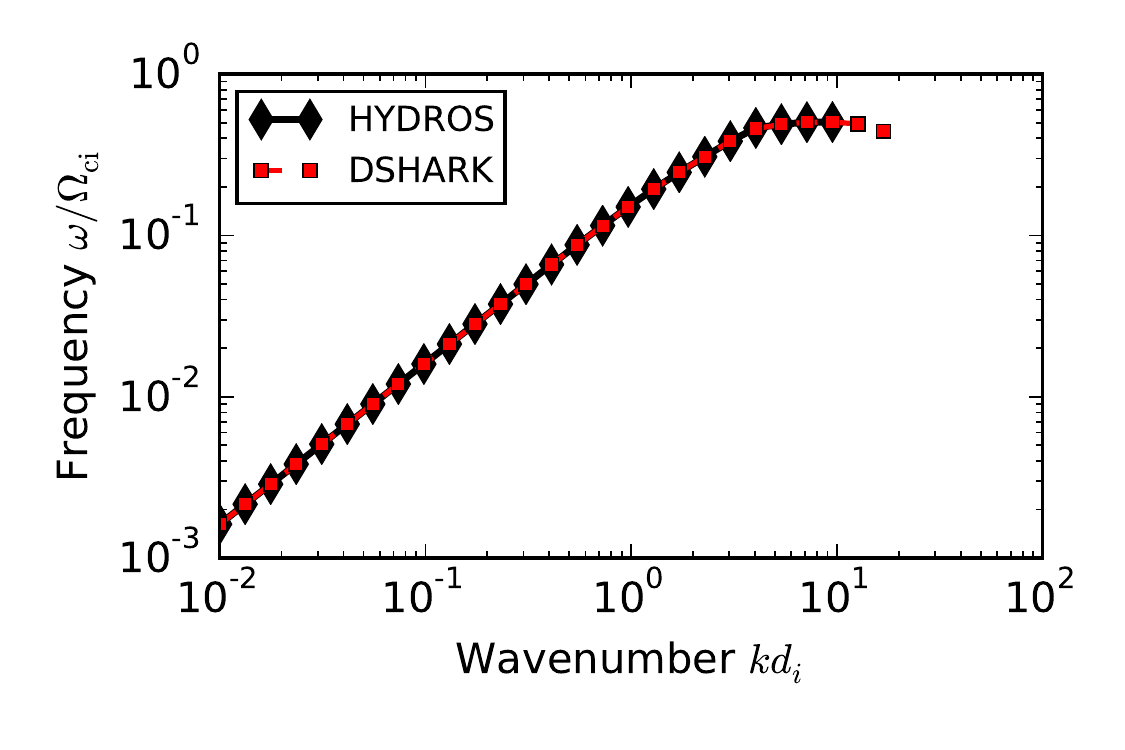}

}\subfloat[\label{fig:SMg-1}]{\includegraphics[width=0.5\textwidth]{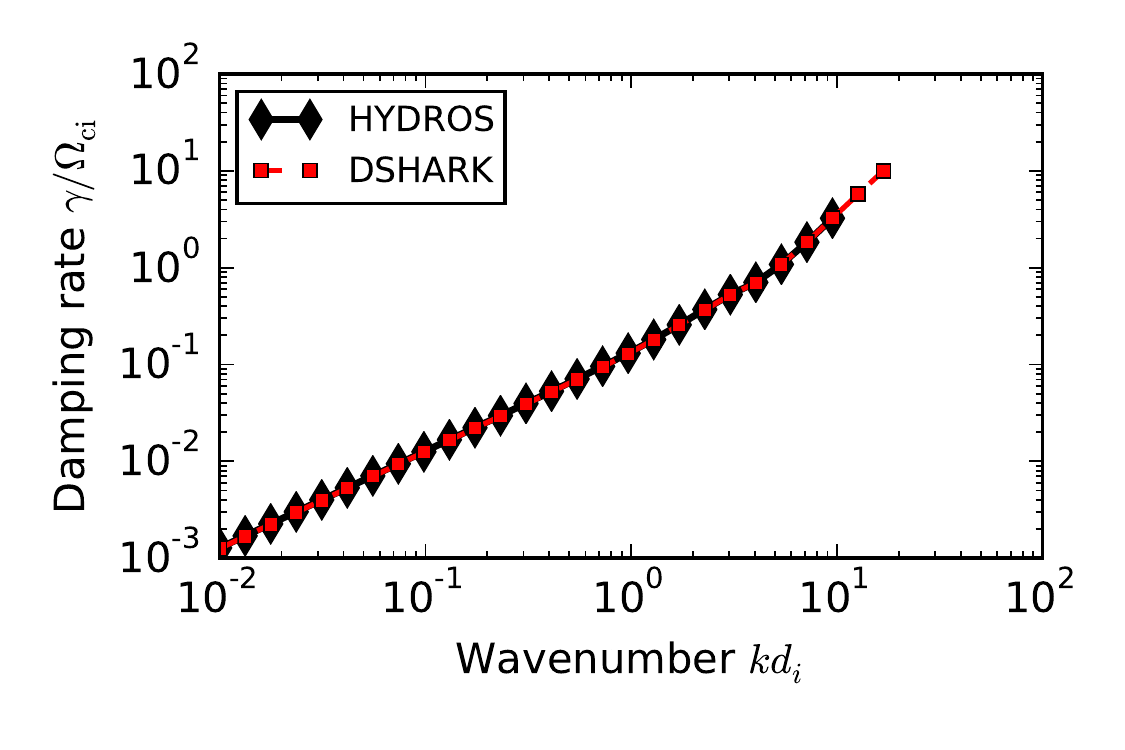}

}

\caption{Comparison of slow magnetosonic mode frequencies (a) and damping rates
(b) between HYDROS and DSHARK. The parameters are $\beta_{\protect\e}=\beta_{\protect\p\|}=\beta_{\protect\p\perp}=1$,
and a propagation angle of $85^{\circ}$.}
\end{figure}

Finally, we study the dispersion relation of the slow magnetosonic
mode as we transition from the MHD spatial scales into the kinetic
range. As before, we use $\beta_{\i\|}=\beta_{\i\perp}=\beta_{\e}=1$,
and a propagation angle of 85$^{\circ}$. We start the scans from
$kd_{\i}=0.01$, and both solvers are able to track this heavily damped
mode down to $kd_{\i}\gtrsim10$, where its damping rate exceeds the
frequency by about a factor 10. In this case, no adjustment of the
electron mass is required, and the HYDROS results agree very well
with those of DSHARK, as is demonstrated in Figs.~\ref{fig:SMw}
and \ref{fig:SMg-1}.

\section{Conclusions}

In this work, we have derived a dispersion relation for a widely used
hybrid kinetic-ion/fluid-electron model of plasma physics, which comprises
the full physics content of a kinetic ion description, while retaining
only a simple massless electron fluid model. This dispersion relation
is valid for arbitrary propagation angle, and retains basic anisotropy
effects by allowing for a bi-Maxwellian background ion distribution,
while focusing on the description of a single ion species. 

We have described the implementation of the dispersion solver HYDROS,
which utilizes the facilities of the Python/SciPy language to solve
the hybrid dispersion relation numerically. The correctness of our
derivation and implementation has been verified against the fully
kinetic code DSHARK, for a variety of waves and instabilities that
propagate parallel, perpendicularly, or obliquely with respect to
the background magnetic field. 

In all examined cases, the hybrid-kinetic dispersion solver showed
the expected behavior and faithfully reproduced the ion physics contained
also in a fully kinetic model. In some instances, it was necessary
to numerically approach the massless electron limit in the fully kinetic
code in order to obtain good agreement, indicating the impact of electron
physics that is not captured in the present version of the hybrid-kinetic
model. Although it was always possible in principle to recover the
hybrid results by simply using very light electrons in DSHARK, the
use of a dedicated hybrid solver proved to be numerically advantageous,
since the occurrence of diverging Bessel arguments is avoided. Furthermore,
HYDROS is suitable to serve as a testbed for additional linear physics
introduced through more sophisticated electron fluid models, which
are not obtained by taking the small mass ratio limit of full kinetics. 

We emphasize that in the present paper we made no attempt to examine
the quality or applicability of the hybrid model itself for real-world
plasmas. Instead, our main purpose here was to demonstrate the correctness
of both our derivation and its numerical implementation. In future
work, we aim to fill this gap and give a more complete account of
the physics fidelity of the hybrid model for both natural and laboratory
plasmas, and its relative merit in comparison to other reduced models
such as gyrokinetic theory. In particular, we expect such a study
to have a bearing on turbulence studies for solar wind and magnetospheric
plasmas, for which hybrid-kinetic simulation codes are a common workhorse.
To this end, the HYDROS code has been made publicly available in a
Github repository \cite{Hydros}. 

\label{sec:conclusions}

\section*{Acknowledgments}

J. Cookmeyer gratefully acknowledges support from NSF Grant PHY-1460055
through the UCLA Physics \& Astronomy REU program. The research leading
to these results has received funding from the European Research Council
under the European Union's Seventh Framework Programme (FP7/2007-2013)/ERC
Grant Agreement No. 277870. Furthermore, this work was facilitated
by the Plasma Science and Technology Institute at UCLA, and by the
Max-Planck/Princeton Center for Plasma Physics.

\section*{Appendix: Gyroangle integrals}

\label{sec:Integrals}In this appendix, we describe the solution method
for the integrals that were introduced in Eq.~(\ref{eq:integrals})
of Sec.~\ref{sec:deriv}. Their definition, written formally as 
\[
I_{abc}=\int_{\mathcal{V}}\int_{-\sigma\oo}^{\theta}\mathcal{K}\left(\theta,\theta'\right)\vpa^{a}\F f(\theta)h(\theta')d\theta'd^{3}v,
\]
with 
\[
\mathcal{K}\left(\theta,\theta'\right)=\exp\left(\im\nu\left(\theta'-\theta\right)+\im\kappa\left(\sin\theta-\sin\theta'\right)\right),
\]
 results in 18 separate integrals. In order to solve the dispersion
relation for the hybrid model equations as introduced in Sec.~\ref{sec:Eq},
we require only those 15 integrals (the ones defined by $a=0\land c=0$
do not appear), which are explicitly written as 
\begin{eqnarray*}
I_{001} & = & \int_{0}^{\oo}\int_{-\oo}^{\oo}\int_{0}^{2\pi}\int_{-\sigma\oo}^{\theta}\mathcal{K}\left(\theta,\theta'\right)v_{\perp}^{2}\F\cos\theta'd\theta'd\theta dv_{\|}d\vp\\
I_{002} & = & \int_{0}^{\oo}\int_{-\oo}^{\oo}\int_{0}^{2\pi}\int_{-\sigma\oo}^{\theta}\mathcal{K}\left(\theta,\theta'\right)v_{\perp}^{2}\F\sin\theta'd\theta'd\theta dv_{\|}d\vp\\
I_{011} & = & \int_{0}^{\oo}\int_{-\oo}^{\oo}\int_{0}^{2\pi}\int_{-\sigma\oo}^{\theta}\mathcal{K}\left(\theta,\theta'\right)v_{\perp}^{3}\cos\theta\F\cos\theta'd\theta'd\theta dv_{\|}d\vp\\
I_{012} & = & \int_{0}^{\oo}\int_{-\oo}^{\oo}\int_{0}^{2\pi}\int_{-\sigma\oo}^{\theta}\mathcal{K}\left(\theta,\theta'\right)v_{\perp}^{3}\cos\theta\F\sin\theta'd\theta'd\theta dv_{\|}d\vp\\
I_{021} & = & \int_{0}^{\oo}\int_{-\oo}^{\oo}\int_{0}^{2\pi}\int_{-\sigma\oo}^{\theta}\mathcal{K}\left(\theta,\theta'\right)v_{\perp}^{3}\sin\theta\F\cos\theta'd\theta'd\theta dv_{\|}d\vp
\end{eqnarray*}
\begin{eqnarray*}
I_{022} & = & \int_{0}^{\oo}\int_{-\oo}^{\oo}\int_{0}^{2\pi}\int_{-\sigma\oo}^{\theta}\mathcal{K}\left(\theta,\theta'\right)v_{\perp}^{3}\sin\theta\F\sin\theta'd\theta'd\theta dv_{\|}d\vp\\
I_{100} & = & \int_{0}^{\oo}\int_{-\oo}^{\oo}\int_{0}^{2\pi}\int_{-\sigma\oo}^{\theta}\mathcal{K}\left(\theta,\theta'\right)v_{\|}\vp\F d\theta'd\theta dv_{\|}d\vp\\
I_{101} & = & \int_{0}^{\oo}\int_{-\oo}^{\oo}\int_{0}^{2\pi}\int_{-\sigma\oo}^{\theta}\mathcal{K}\left(\theta,\theta'\right)v_{\|}v_{\perp}^{2}\F\cos\theta'd\theta'd\theta dv_{\|}d\vp\\
I_{102} & = & \int_{0}^{\oo}\int_{-\oo}^{\oo}\int_{0}^{2\pi}\int_{-\sigma\oo}^{\theta}\mathcal{K}\left(\theta,\theta'\right)v_{\|}v_{\perp}^{2}\F\sin\theta'd\theta'd\theta dv_{\|}d\vp\\
I_{110} & = & \int_{0}^{\oo}\int_{-\oo}^{\oo}\int_{0}^{2\pi}\int_{-\sigma\oo}^{\theta}\mathcal{K}\left(\theta,\theta'\right)v_{\|}v_{\perp}^{2}\cos\theta\F d\theta'd\theta dv_{\|}d\vp
\end{eqnarray*}
\begin{eqnarray*}
I_{111} & = & \int_{0}^{\oo}\int_{-\oo}^{\oo}\int_{0}^{2\pi}\int_{-\sigma\oo}^{\theta}\mathcal{K}\left(\theta,\theta'\right)v_{\|}v_{\perp}^{3}\cos\theta\F\cos\theta'd\theta'd\theta dv_{\|}d\vp\\
I_{112} & = & \int_{0}^{\oo}\int_{-\oo}^{\oo}\int_{0}^{2\pi}\int_{-\sigma\oo}^{\theta}\mathcal{K}\left(\theta,\theta'\right)v_{\|}v_{\perp}^{3}\cos\theta\F\sin\theta'd\theta'd\theta dv_{\|}d\vp\\
I_{120} & = & \int_{0}^{\oo}\int_{-\oo}^{\oo}\int_{0}^{2\pi}\int_{-\sigma\oo}^{\theta}\mathcal{K}\left(\theta,\theta'\right)v_{\|}v_{\perp}^{2}\sin\theta\F d\theta'd\theta dv_{\|}d\vp\\
I_{121} & = & \int_{0}^{\oo}\int_{-\oo}^{\oo}\int_{0}^{2\pi}\int_{-\sigma\oo}^{\theta}\mathcal{K}\left(\theta,\theta'\right)v_{\|}v_{\perp}^{3}\sin\theta\F\cos\theta'd\theta'd\theta dv_{\|}d\vp\\
I_{122} & = & \int_{0}^{\oo}\int_{-\oo}^{\oo}\int_{0}^{2\pi}\int_{-\sigma\oo}^{\theta}\mathcal{K}\left(\theta,\theta'\right)v_{\|}v_{\perp}^{3}\sin\theta\F\sin\theta'd\theta'd\theta dv_{\|}d\vp
\end{eqnarray*}
All integrals can be calculated with the same approach, which will
be demonstrated here for the first integral
\[
I_{001}=\int_{0}^{\oo}\int_{-\oo}^{\oo}\int_{0}^{2\pi}\int_{-\sigma\oo}^{\theta}\mathcal{K}\left(\theta,\theta'\right)v_{\perp}^{2}\F\cos\theta'd\theta'd\theta dv_{\|}d\vp.
\]
As a first step, we solve the innermost integral over $\theta'$ by
means of substitution. We set $\theta'=\theta-\sigma\phi$ and obtain
\[
\fl I_{001}=\sigma\int_{0}^{\oo}\int_{-\oo}^{\oo}\int_{0}^{2\pi}\int_{0}^{\oo}\exp\left(-\im\nu\sigma\phi+\im\kappa\left(\sin\theta-\sin\left(\theta-\sigma\phi\right)\right)\right)v_{\perp}^{2}\F\cos\left(\theta-\sigma\phi\right)d\phi d\theta dv_{\|}d\vp.
\]
Next, we write $\cos\left(\theta-\sigma\phi\right)$ in exponential
form and separate the integrand into a secular part and a periodic
part $h\left(\phi\right)$. Then we replace the infinite $\phi$ integration
boundaries by an infinite sum of $2\pi$ slices, i.e. 
\[
\int_{0}^{\infty}F\left(\phi\right)d\phi\rightarrow\int_{2\pi n}^{2\pi(n+1)}\sum_{n=0}^{\infty}F\left(\phi\right)d\phi.
\]
Applied to the integrand of $I_{001}$ this reads 
\[
I_{001}=\sigma\int_{0}^{\oo}\int_{-\oo}^{\oo}\int_{0}^{2\pi}\sum_{n=0}^{\oo}\int_{2\pi n}^{2\pi\left(n+1\right)}\exp\left(-\im\nu\sigma\phi\right)h\left(\phi\right)d\phi d\theta v_{\perp}^{2}\F dv_{\|}d\vp.
\]
We introduce a shift $\phi'=\phi-2\pi n$:
\begin{eqnarray*}
\fl I_{001}=\sigma\int_{0}^{\oo}\int_{-\oo}^{\oo}\int_{0}^{2\pi}\sum_{n=0}^{\oo}\int_{0}^{2\pi}\exp\left(-\im\nu\sigma\phi'-\im\nu\sigma2\pi n\right)h\left(\phi'\right)d\phi'd\theta v_{\perp}^{2}\F dv_{\|}d\vp\\
=\sigma\int_{0}^{\oo}\int_{-\oo}^{\oo}\int_{0}^{2\pi}\sum_{n=0}^{\oo}\exp\left(-\im\nu\sigma2\pi n\right)\int_{0}^{2\pi}\exp\left(-\im\nu\sigma\phi'\right)h\left(\phi'\right)d\phi'd\theta v_{\perp}^{2}\F dv_{\|}d\vp\\
=Q\left(\nu\right)\sigma\int_{0}^{\oo}\int_{-\oo}^{\oo}\int_{0}^{2\pi}\int_{0}^{2\pi}\exp\left(-\im\nu\sigma\phi'\right)h\left(\phi'\right)d\phi'd\theta v_{\perp}^{2}\F dv_{\|}d\vp
\end{eqnarray*}
where 
\[
Q\left(\nu\right)=\sum_{n=0}^{\oo}\exp\left(-\im\nu\sigma2\pi n\right).
\]
For damped modes, we have $\sigma=1$ and $\Im\left(\nu\right)<0$,
while for unstable modes $\Im\left(\nu\right)>0$ and $\sigma=-1$.
Because of the choices made for the integration boundaries and the
substitution of $\theta'$, the geometric series 
\[
Q\left(\nu\right)=\sum_{n=0}^{\oo}\exp\left(-\im\nu\sigma2\pi n\right)=\frac{1}{1-\exp\left(-\im\nu\sigma2\pi\right)}
\]
converges in both cases. Next, we insert the definition of the Bessel
functions
\[
\exp\left(\pm\im z\sin x\right)=\sum_{n=-\oo}^{\oo}\exp\left(\pm\im nx\right)J_{n}\left(z\right)
\]
 and apply some simple manipulations to arrive at
\begin{eqnarray*}
\fl I_{001}=\frac{1}{2}Q\left(\nu\right)\sigma\int_{0}^{\oo}\int_{-\oo}^{\oo}\int_{0}^{2\pi}\int_{0}^{2\pi}\exp\left(-\im\nu\sigma\phi\right)J_{n}\left(\kappa\right)J_{m}\left(\kappa\right)v_{\perp}^{2}\F\\
\times\sum_{n=-\oo}^{\oo}\sum_{m=-\oo}^{\oo}\Bigl(\mathrm{e}^{\im\left(n-m+1\right)\theta}\mathrm{e}^{\im\sigma\left(m-1\right)\phi}+\mathrm{e}^{\im\left(n-m-1\right)\theta}\mathrm{e}^{\im\sigma\left(m+1\right)\phi}\Bigr)d\theta d\phi dv_{\|}d\vp.
\end{eqnarray*}
Next, we may perform the integration over $\theta$, which yields
\begin{eqnarray*}
\fl I_{001}=\pi Q\left(\nu\right)\sigma\int_{0}^{\oo}\int_{-\oo}^{\oo}\int_{0}^{2\pi}\sum_{m=-\oo}^{\oo}\Bigl(\mathrm{e}^{\im\sigma\left(m-\nu-1\right)\phi}J_{m-1}\left(\kappa\right)J_{m}\left(\kappa\right)\\
+\mathrm{e}^{\im\sigma\left(m-\nu+1\right)\phi}J_{m+1}\left(\kappa\right)J_{m}\left(\kappa\right)\Bigr)d\phi v_{\perp}^{2}\F dv_{\|}d\vp.
\end{eqnarray*}
 Upon performing the $\phi$ integration, the geometric series $Q\left(\nu\right)$
as well as the factor $\sigma$ cancel out and we obtain
\begin{eqnarray*}
\fl I_{001}=\im\pi\int_{0}^{\oo}\int_{-\oo}^{\oo}\sum_{m=-\oo}^{\oo}\biggl(\frac{1}{m-\nu-1}J_{m-1}\left(\kappa\right)J_{m}\left(\kappa\right)+\frac{1}{m-\nu+1}J_{m+1}\left(\kappa\right)J_{m}\left(\kappa\right)\biggr)v_{\perp}^{2}\F dv_{\|}d\vp.\\
\end{eqnarray*}
 Now, it remains to perform the integrations in $\vpa$ and $\vp$.
We write 
\[
\nu=\frac{\omega}{\Omega_{c}}-\frac{\kp v_{\|}}{\Omega_{c}}=\frac{1}{\xi_{c}}\left(\xi-x\right)
\]
with 
\[
\xi=\frac{\omega}{\kp\vtpa},\qquad\xi_{c}=\frac{\Omega_{c}}{\kp\vtpa},\qquad x=\frac{\vpa}{\vtpa}
\]
and insert $\F$ from Eq.~(\ref{eq:F0}) to obtain 
\begin{eqnarray*}
\fl I_{001}=\im\pi\frac{\m n_{0\i}}{2\pi}\left(T_{0\i\perp}^{2}\right)^{-1/2}\int_{0}^{\oo}\exp\left(-\frac{v_{\perp}^{2}}{v_{\mathrm{th}\perp i}^{2}}\right)\sum_{m=-\oo}^{\oo}\Bigl(\xi_{c}Z\left(\xi-\xi_{c}\left(m-1\right)\right)J_{m-1}\left(\kappa\right)J_{m}\left(\kappa\right)\\
+\xi_{c}Z\left(\xi-\xi_{c}\left(m+1\right)\right)J_{m+1}\left(\kappa\right)J_{m}\left(\kappa\right)\Bigr)v_{\perp}^{2}d\vp.\\
\end{eqnarray*}
Here, we have introduced the plasma dispersion function \cite{Fried61}
\[
Z\left(\xi\right)=\frac{1}{\sqrt{\pi}}\int_{\oo}^{\oo}dx\frac{\exp\left(-x^{2}\right)}{x-\xi}.
\]
Finally, we perform the $\vp$ integration and, introducing a further
abbreviation 
\[
\mathcal{C}=\frac{n_{0i}\Omega_{c}}{\kp\vtpa},
\]
 obtain the solution
\[
\fl I_{001}=\im\mathcal{C}\frac{\Omega_{c}}{2\kx}\sum_{m=-\oo}^{\oo}\biggl(m\left(Y_{m-1}+Y_{m+1}\right)\Gamma_{m}\left(\zeta\right)+\frac{\kx^{2}\vtp^{2}}{2\Omega_{c}^{2}}\left(Y_{m-1}-Y_{m+1}\right)\Gamma_{m}'\left(\zeta\right)\biggr).
\]
We have introduced the abbreviations 
\begin{eqnarray*}
Y_{m} & = & Z\left(\xi-\xi_{c}m\right)
\end{eqnarray*}
and
\[
X_{m}=1+\left(\xi-\xi_{c}m\right)Z\left(\xi-\xi_{c}m\right),
\]
the latter of which appears when performing these steps for integrals
with $a=1$. Also, we have introduced the exponentially scaled modified
Bessel functions
\[
\Gamma_{m}\left(\zeta\right)=I_{m}\left(\zeta\right)\exp\left(-\zeta\right),
\]
with their argument 
\[
\zeta=\frac{1}{2}\left(\frac{\kx\vtp}{\Omega_{c}}\right)^{2}.
\]
Since the $\Gamma_{m}$ function and its derivative $\Gamma_{m}'$
appear exclusively with the argument $\zeta$, it is omitted in the
following. The remaining integrals can be solved in the same fashion,
and their final expressions read

\[
\fl I_{001}=\im\mathcal{C}\frac{\Omega_{c}}{2\kx}\sum_{m=-\oo}^{\oo}\left[m\left(Y_{m-1}+Y_{m+1}\right)\Gamma_{m}+\zeta\left(Y_{m-1}-Y_{m+1}\right)\Gamma_{m}'\right]
\]
\[
\fl I_{002}=\mathcal{C}\frac{\Omega_{c}}{2\kx}\sum_{m=-\oo}^{\oo}\left[m\left(Y_{m-1}-Y_{m+1}\right)\Gamma_{m}+\zeta\left(Y_{m-1}+Y_{m+1}\right)\Gamma_{m}'\right]
\]
\begin{eqnarray*}
\fl I_{011}=\frac{\im}{2}\mathcal{C}\vtp^{2}\sum_{m=-\oo}^{\oo}\left(m-1\right)Y_{m-1}\Gamma_{m}'+\im\mathcal{C}\left(\frac{\Omega_{c}}{\kx}\right)^{2}\sum_{m=-\oo}^{\oo}m\left(m-1\right)Y_{m-1}\Gamma_{m}\\
+\frac{\im}{4}\mathcal{C}\vtp^{2}\sum_{m=-\oo}^{\oo}\left(Y_{m+1}-Y_{m-1}\right)\left(\zeta\Gamma_{m}'+\Gamma_{m}\right)
\end{eqnarray*}
\[
\fl I_{012}=-I_{021}
\]
\[
\fl I_{021}=\frac{1}{4}\mathcal{C}\vtp^{2}\sum_{m=-\oo}^{\oo}\left(Y_{m+1}-Y_{m-1}\right)\left(\zeta\Gamma_{m}'+\Gamma_{m}\right)
\]
\begin{eqnarray*}
\fl I_{022}=-\frac{\im}{2}\mathcal{C}\vtp^{2}\sum_{m=-\oo}^{\oo}\left(m-1\right)Y_{m-1}\Gamma_{m}'-\im\mathcal{C}\left(\frac{\Omega_{c}}{\kx}\right)^{2}\sum_{m=-\oo}^{\oo}m\left(m-1\right)Y_{m-1}\Gamma_{m}\\
+\frac{\im}{4}\mathcal{C}\vtp^{2}\sum_{m=-\oo}^{\oo}\left(Y_{m+1}+3Y_{m-1}\right)\left(\zeta\Gamma_{m}'+\Gamma_{m}\right)
\end{eqnarray*}
\begin{eqnarray*}
\fl I_{100} & = & \im\mathcal{C}\vtpa\sum_{m=-\oo}^{\oo}X_{m}\Gamma_{m}
\end{eqnarray*}
\begin{eqnarray*}
\fl I_{101}=\im\mathcal{C}\vtpa\frac{\Omega_{c}}{2\kx}\sum_{m=-\oo}^{\oo}\left[m\left(X_{m-1}+X_{m+1}\right)\Gamma_{m}+\zeta\left(X_{m-1}-X_{m+1}\right)\Gamma_{m}'\right]
\end{eqnarray*}
\begin{eqnarray*}
\fl I_{102}=\mathcal{C}\vtpa\frac{\Omega_{c}}{2\kx}\sum_{m=-\oo}^{\oo}\left[m\left(X_{m-1}-X_{m+1}\right)\Gamma_{m}+\zeta\left(X_{m-1}+X_{m+1}\right)\Gamma_{m}'\right]
\end{eqnarray*}
\[
\fl I_{110}=I_{101}
\]
\begin{eqnarray*}
\fl I_{111}=\frac{\im}{2}\mathcal{C}\vtpa\vtp^{2}\sum_{m=-\oo}^{\oo}\left(m-1\right)X_{m-1}\Gamma_{m}'+\im\mathcal{C}\vtpa\left(\frac{\Omega_{c}}{\kx}\right)^{2}\sum_{m=-\oo}^{\oo}m\left(m-1\right)X_{m-1}\Gamma_{m}\\
+\frac{\im}{4}\mathcal{C}\vtpa\vtp^{2}\sum_{m=-\oo}^{\oo}\left(X_{m+1}-X_{m-1}\right)\left(\zeta\Gamma_{m}'+\Gamma_{m}\right)
\end{eqnarray*}
\[
\fl I_{112}=-I_{121}
\]
\[
\fl I_{120}=-I_{102}
\]
\[
\fl I_{121}=\frac{1}{4}\mathcal{C}\vtpa\vtp^{2}\sum_{m=-\oo}^{\oo}\left(X_{m+1}-X_{m-1}\right)\left(\zeta\Gamma_{m}'+\Gamma_{m}\right)
\]
\begin{eqnarray*}
\fl I_{122}=-\frac{\im}{2}\mathcal{C}\vtpa\vtp^{2}\sum_{m=-\oo}^{\oo}\left(m-1\right)X_{m-1}\Gamma_{m}'-\im\mathcal{C}\vtpa\left(\frac{\Omega_{c}}{\kx}\right)^{2}\sum_{m=-\oo}^{\oo}m\left(m-1\right)X_{m-1}\Gamma_{m}\\
+\frac{\im}{4}\mathcal{C}\vtpa\vtp^{2}\sum_{m=-\oo}^{\oo}\left(X_{m+1}+3X_{m-1}\right)\left(\zeta\Gamma_{m}'+\Gamma_{m}\right).
\end{eqnarray*}
These results can be inserted into the matrix elements of Eq.~(\ref{eq:DR}),
enabling a numerical solution of the dispersion relation.


\end{document}